\documentclass{aastex}

\newcommand{\Rsun}{\ensuremath{R_{\odot}}}
\newcommand{\Msun}{\ensuremath{M_{\odot}}}

\def \ie{{\sl i.e.}}
\def \eg{{\sl e.g.}}
\def \viz{{\sl viz.}}

\newcommand{\insertfig}[4]{
  \begin{figure}[!htbp]
   \begin{center}
      \includegraphics[#1]{#2}
   \end{center}
   \vspace{-0.00cm}
   \caption{#3}
   \label{#4}
  \end{figure}
}

\newcommand{\insertdoblfig}[6]{
  \begin{figure}[!htbp]
   \begin{center}
     \includegraphics[#1]{#2}
     \includegraphics[#3]{#4}
   \end{center}
   \vspace{-0.00cm}
   \caption{#5}
   \label{#6}
  \end{figure}
}

\usepackage{graphicx,natbib,subfigure}

\title{Collective Properties of X-ray Binary Populations of Galaxies II.
  \\ Pre-Low-mass X-ray Binary Properties, Formation Rates, and Constraints}

\author{H. Bhadkamkar\altaffilmark{1, 2} and P. Ghosh\altaffilmark{1} }

\altaffiltext{1}{Department of Astronomy \& Astrophysics, Tata 
	Institute of Fundamental Research, Mumbai 400 005, India}

\altaffiltext{2}{Astronomy \& Astrophysics, Raman Research Institute,
	Bengaluru, 560080}

\begin{abstract}
We continue exploring our understanding of the collective
properties of X-ray binaries in the stellar fields (\ie, 
outside globular clusters) of normal galaxies, introduced in 
Paper I of this series, where we considered high-mass X-ray 
binaries (HMXBs). In this paper (Paper II of the series) and 
the companion paper (Paper III of the series), we consider 
low-mass X-ray binaries (LMXBs), whose evolutionary scenario 
is very different from that of HMXBs. In this paper, we
consider the evolution of primordial binaries upto the stage
where the neutron star just formed in the supernova explosion 
of the primary is in a binary with its low-mass unevolved 
companion, and this binary has circularized tidally, producing
what we call a pre-low-mass X-ray binary (pre-LMXB). We study
the constraints on the formation of such pre-LMXBs in detail
(since these are low-probability events), and calculate their
collective properties and formation rate. To this end, we 
first consider the changes in the binary parameters in the
various steps involved, \viz, the common-envelope (CE) phase,
the supernova, and the tidal evolution. This naturally leads
to a clarification of the constraints. We then describe our
calculation of the evolution of the distributions of primordial
binary parameters into those of the pre-LMXB parameters,
following the standard evolutionary scenario for individual
bianries. We display the latter as both bivariate and 
monovariate distributions, discuss their essential properties,
and indicate the influence of some essential factors on these.
Finally, we calculate the formation rate of pre-LMXBs. The
results of this paper are used in the next one (Paper III) to
compute the expected X-ray luminosity function (XLF) of LMXBs,
which is compared with observation.                  
\end{abstract}


\begin{document}
\maketitle

\section{Introduction}

In this series of papers, we are attempting a pioneering exploration
of the essential theoretical underpinnings of the observed
distributions of the \emph{collective} properties of accretion 
powered X-ray binaries, in particular their X-ray luminosity functions
(XLFs). Paper I \citep{hhmxb} of the series dealt with high-mass (or massive)
X-ray binaries (HMXBs). In the next two papers of the series, we are
dealing with low-mass X-ray binaries (LMXBs). It is convenient to
divide this LMXB study into two parts, \viz, (a) that which deals with
the evolution of a primordial binary upto the formation of a neutron
star in a supernova (SN) explosion, which produces a binary consisting
of this neutron star and a relatively low-mass companion, which we 
call a pre-low-mass X-ray binary (pre-LMXB), and (b) that which deals
with the evolution of this pre-LMXB into an X-ray active LMXB, and 
the subsequent evolution of this LMXB through its accretion phase.
Accordingly, we handle the first part in this Paper, which is Paper II
in this series. The second part is handled in a companion paper, which
is Paper III in this series.

We briefly recount a few essential features of studies such this.
First, such studies have become meaningful only in recent years,
when it became possible to construct robust and dependable 
distributions of the essential collective properties of X-ray
binaries, \eg, their XLFs, after the accumulation of four decades
of observational material \citep{grimm01, grimm03, kim, gilfanov04,
gg46, gg41, liulmxb, kimfab10}. Second, in our study contained in this
series of papers, we focus on X-ray binaries which are \emph{outside}
globular clusters, \ie, in the stellar field of the galaxy under
study, which implies that the evolution of a given X-ray binary can
be treated in isolation, without any significant perturbation from
other stars or X-ray binaries. This is why the standard scenarios of
individual X-ray binary evolution \citep{heuvelrev83, heuvelrev91, 
heuvelrev92, heuvelrev01} can be applied to the problem. Third, the multi-step 
evolutionary sequence from a primordial binary to an X-ray binary
may involve both (a) steps in which only the initial and final
states matter for our purposes, and (b) steps in which not only
the initial and final states but also the entire intermediate 
process of evolution have to be considered for our purposes.
Naturally, the second situation leads to more involved calculations.
We showed in Paper I that the entire HMXB evolutionary process is
described by the first situation. We shall show in this paper 
(Paper II) that the evolution from primordial X-ray binaries to 
pre-LMXBs is also almost described by the first situation, with 
one exception which is easily handled. By contrast, the subsequent 
LMXB evolution is almost entirely described by the second situation, 
and so has to be handled in a very different manner. This is a major 
reason why we chose to give it in a separate paper, \viz, Paper III. 
Fourth, as with the HMXB work of Paper I, an effort of this type 
should be regarded \emph{only} as a proof-of-principle demonstration 
that the observed X-ray-binary collective properties can be (at 
least) qualitatively accounted for by evolving well-known, plausible
collective properties of primordial binaries through the well-known
and well-accepted scenarios for the evolution of an \emph{individual} 
primordial binary into an X-ray binary.      
      
The rest of this paper is arranged as follows. In Sec.\ref{scenario},
we briefly recount the formation and evolution scenarios of 
pre-LMXBs and LMXBs. In Sec.\ref{paramchange}, we describe the 
changes in the binary parameters in various steps of primordial-binary
evolution leading upto pre-LMXB formation, taking in turn the 
common-envelope (CE) phase, the supernova, and the tidal evolution.
In Sec.\ref{primoconstr}, we describe the constraints on the 
pre-LMXB parameters. In Sec.\ref{distr}, we describe our calculation 
of the distribution of pre-LMXB parameters. We start from a summary
of canonical primordial parameter distribution, and we describe how
we transform this distribution to obtain that for pre-LMXBs. We 
show the latter as both bivariate and monovariate distributions,
discuss their properties, and show the influence of some essential
factors on these. In Sec.\ref{frate}, we present our calculation of
the formation rate of pre-LMXBs. We discuss our results in 
Sec.\ref{discuss}.     

\section{LMXB Formation and Evolution Scenarios}
\label{scenario}

We briefly recount here the standard formation scenario for LMXBs. Their 
progenitors are primordial binaries of two main sequence stars,
which are much more disparate in mass than those in HMXB progeniors. 
This extreme mass ratio changes the course of LMXB formation and  
evolution completely from that of HMXBs. The more massive star, \ie, 
the {\it primary}, completes its main sequence life faster, on
timescales $\sim 10^6-10^7$ yr. On the giant branch, it fills its 
Roche lobe and starts transferring mass to the secondary, which, 
in such a progenitor, is so much less massive than the primary,
and hence has so much longer a thermal timescale than the primary, 
that it is unable to accept the transferred matter, which then forms
an envelope around the two stars - the common envelope (CE). 
The primary keeps losing mass to form this CE until it is completely stripped
of its H-envelope. The CE, which engulfs the resulting binary of the  
He-core of the primary and the secondary, exerts a strong frictional 
drag on this binary, so that the two stars spiral in towards each
other. The orbital energy released due to this spiral-in is deposited 
into the CE, heats it and attempts to expel it. Systems which were very
close to begin with now go into a merger and so do not survive this
phase, but those which were sufficiently wide do have enough energy 
to expel the CE altogether, and so survive the CE phase, emerging as
a very compact binary consisting of the He-core of the primary and the 
secondary, the latter remaining practically unchanged during this 
whole process.

The post-CE system is typically detached and the He-core evolves as if
it were a single He-star, exploding eventually as a supernova (SN). 
A neutron star of typical mass 1.4 \Msun\ is left after the SN explosion of
the He-core, the rest of the mass being lost from the system. Due to 
the large mass loss from the system during the previous CE phase, the
mass loss in this SN event is not always destructively large, and some of these
systems do survive, but often with rather large eccentricity. The rest of
the systems are disrupted. Natal kicks are given to the neutron star
when the SN explosion is asymmetric, and these further influence the
survival probability of these post-CE binaries. The resultant binary,
consisting of the neutron star and its low-mass companion, now evolves 
tidally on timescales $\sim 10^5-10^6$ yr, circularizing the binary.  
The whole sequence of events outlined above occurs over $\sim 10^7$
yr, which is very short compared to the nuclear timescale of the 
secondary, which for a typical secondary mass of 1\Msun\ is $>10^9$
years. Therefore, the secondary undergoes little nuclear evolution 
during the above time, and can still be considered almost a 
Zero Age Main Sequence (ZAMS) star. We call this binary system of a 
neutron star and low-mass unevolved companion a \emph{Pre-Low-Mass
X-ray Binary (pre-LMXB)}.

LMXB formation and evolution can thus be thought of as a
two-step process. The first step is the rapid one outlined above,
occurring over timescales of $\sim 10^7$ yr. The second step is a
very slow one, occurring over timescales $\sim 10^9$ years, during
which (a) the above pre-LMXB evolves through its detached binary phase to 
the onset of Roche-lobe overflow and mass transfer, at which point
the system turns on as a LMXB, and (b) this LMXB lives out
its X-ray emission phase and turns off, after which system evolves
into a binary consisting of a recycled neutron star with a white
dwarf companion. We treat the first step in detail in the rest of 
this paper (Paper II), deferring a detailed treatment of the second 
step to the next paper of this series (Paper III), and giving only 
a very brief outline of that step in the next paragraph.  

The pre-LMXB evolves through the angular momentum loss which causes 
its orbit to shrink, the two major mechanisms for such loss being 
gravitational radiation and magnetic braking. The first is always in 
operation, though not always dominant. Magnetic braking is operational 
when the companion has a sufficiently large and sustained magnetic
field, and a magnetically coupled stellar wind carries off 
significant angular momentum. This is thought to require both a
sizable radiative core (to anchor the magnetic field) and a
sizable convective envelope (to run the dynamo that produces the 
magnetic field) in the companion, a point to which we return later. 
Both mechanisms have strengths which increase rapidly with decreasing
orbital radii, and so are important only at small radii. Accordingly,
these mechanisms can bring the system into Roche-lobe contact within
the main-sequence lifetime of the low-mass companion only if the 
initial orbital radius of the pre-LMXB is sufficiently small. For
such systems, Roche-lobe overflow is established while the companion 
is still on the main-sequence, and the subsequent mass transfer is
driven by angular-momentum loss. For wider systems, however, 
Roche-lobe contact becomes possible only after the companion finishes
its main-sequence lifetime, gets on the giant branch and expands to
fill its Roche lobe. Subsequent mass transfer is driven dominantly by
the nuclear evolution and expansion of the companion. The crossover
point between the above two possible evolutionary paths occurs at a 
critical initial orbital period called the \emph{bifurcation period}, 
whose value is $\sim 14-18$ hours \citep{pylyser88, pylyser89}. 
In either case, the matter transferred to the neutron star's 
Roche-lobe forms an accretion disk and eventually reaches the surface 
of the neutron star, the energy released in the process being emitted 
in X-rays, making the system a bright LMXB. This LMXB phase ends
when accretion stops and the companion becomes a degenerate white dwarf,
the final product being a recycled neutron star in either (a) a 
close orbit with either a low-mass He white dwarf or a somewhat 
heavier CO white dwarf, or (b) a wide orbit with a low-mass He
white dwarf. The former class comes from the angular-momentum-loss
driven evolution of systems below the bifurcation period, while
the latter class comes from the companion-nuclear-evolution driven 
evolution of systems above the bifurcation period.  

We note here that other scenarios for the formation of LMXBs have 
been suggested, \eg, accretion induced collapse of a white dwarf. 
(See,\eg, \citet{kalogera98, kalogera98a} and references therein). 
However, most of the scenarios relevant for the LMXB systems in the 
field of the galaxy invoke CE scenario and hence fall in the general
class whose evolutionary sequence is similar to what we described
above. Further, since we are interested here in the LMXB populations 
\emph{outside} globular clusters, we do not consider those scenarios 
which invoke stellar encounters as the dominant mechanism for the 
formation of these systems.

\section{Evolutionary Changes in Binary Parameters}
\label{paramchange}

\subsection{The common envelope (CE) phase}
\label{ce}

The conditions required for the formation of a CE have been studied in
great detail in the literature (for detailed reviews, 
see, \eg, \citet{taam, webbink}). The essential ingredients 
for the viable formation of a CE are (a) an extreme mass
ratio and (b) an evolved primary. Depending upon the evolutionary
stage of the primary, three regimes of the primary's Roche-lobe overflow, 
namely Cases A, B and C, have been defined and are widely used 
in the literature. The cases of interest to us here usually belong to 
Cases B and C, involving an evolved primary. (Case A occurs when the 
donor is on main sequence, which can happen in very close primordial 
binaries, but this case is not of interest here, since it would lead 
to a merger during spiral-in, as explained earlier). The other 
condition, \ie, that of extreme mass ratio, is usually taken as $q <
0.3$, because in this regime the thermal timescales of the two stars 
are different by an order of magnitude or more.
This regime of $q$-values is completely disjoint with that relevant for
the HMXB progenitors (see Paper I), as expected.
The mass of the secondary remains unchanged during the entire
evolution from the primordial binary to Roch-lobe overflow, as 
explained earlier. The upper limit to this mass is usually taken as 
$1 - 2 \Msun$, although some authors have suggested somewhat higher 
limits \citep{podsiadlowski, pfahl03}, in which case there is 
expected to be considerable mass loss in the initial 
phases of LMXB operation. For typical primary masses in primordial 
binaries, which are in the range $9-30 \Msun$ for LMXBs, the mass 
ratio $q$ thus always satisfies the above criterion of extremeness.

Quantitative descriptions of the binary-parameter changes in the 
CE phase can be given from the general picture of CE evolution 
given above \citep{webbink84}. The mass of the common envelope is 
usually taken as the mass of the primary envelope, as only the 
He-core of the primary is left at 
the end of the CE phase. The energy deposited 
in the CE is equal to the difference in the binding energy of the 
pre-CE and post-CE system, multiplied by an efficiency factor 
$\alpha$. This energy must be equal to the core-envelope binding
energy for the primary, if the enevelope is to be expelled finally. 
Thus, if $E_{b1}$ and $E_{b2}$ are the initial 
and final orbital binding energies and $E_{c-e}$ is the energy of 
core-envelope binding, the energy equation describing CE process 
is $\alpha(E_{b1}-E_{b2}) = E_{c-e}$. Now,
the binding energy $E_b$ can be written as $GM_1M_2/2a$ where the
masses and the distance are taken appropriately. Calculation of 
$E_{c-e}$ requires a knowledge of the detailed structure of the star,
and the result can be parametrized as $E_{c-e} = GM_{p,c}M_{p,e}/
\lambda R$ where $M_{p,c}$ and $M_{p,e}$ are the masses of the core 
and the envelope of the primary respectively and $R$ is the radius 
of the star. Details of stellar structure are contained in the 
parameter $\lambda$. 

Since the radius of the star equals that of its Roche lobe at 
the time of CE formation, the relation between the initial and final 
orbital separations can be written as 
\begin{equation}
\frac{a_{CE}}{a_0} = \frac{M_{p,c} M_s}{M_p}
 \frac{1}{M_s + 2M_{p,e}(\alpha \lambda r_L)^{-1}}.
\label{eqn:cechange}
\end{equation}
The relation between $M_{p,c}$ and $M_p$ is often taken as a power-law 
approximation to the results of detailed stellar-structure calculations,
given by 
\begin{equation}
  M_{p,c}=M_0\,M_p^{1/\xi},
  \label{eqn:mcore}
\end{equation}
where $M_{p,e} = M_p - M_{p,c}$. Here, $r_L$ is the effective Roche
lobe radius of the primary. It depends only upon the mass ratio
$q\equiv M_1/M_2$, and analytical approximations to the numerical 
results are available in the literature. We adopt here the  
widely-used \emph{Eggleton approximation} \citep{eggleton}, which 
is
\begin{equation}
r_L = \frac{0.49 q^{2/3}}{0.6 q^{2/3} + \ln(1+q^{1/3})}.
\label{eqn:roche}
\end{equation}
 
From the above energy equation, it is clear that the orbital shrinkage 
during the CE phase depends upon the product $\alpha\lambda$ of the
two essential parameters of the problem. Consequently, a modeling of 
the CE process alone cannot give independent handles on these two 
parameters, but can only constrain their product. However, many
authors have suggested constraints on the values of
$\alpha$ and $\lambda$, which are based on observations of CVs
and LMXBs, and also from the results of stellar-evolution computations.
From these works it is seen that, although values of $\lambda$ can be
estimated with reasonable confidence for a large range of stellar 
parameters, $\alpha$ is still rather uncertain, with expected values 
$\sim 0.5 - 1$ \citep{dewi, willems, marco, ivanova}. Since
independent constraints are not possible from CE modeling alone, as
in this work, we here work only with the product $\alpha\lambda$, 
calling it the \emph{CE parameter}.

The transformation connecting the parameters of the primordial binary 
to those of the post-CE binary are given by Eqs. \ref{eqn:cechange}
and \ref{eqn:mcore} along with the definition $M_s=q\,M_p$. The
inverse transformation can be derived from these, and the Jacobian 
for that transformation is given by
\begin{equation}
J_{CE} = \frac {\xi}{M_{p,c}} \frac{a_0}{a_{CE}}
\label{eqn:jce}
\end{equation}

\subsection{Supernova}
\label{sn}

The detached post-CE system evolves at first without affecting the binary 
parameters. Then the He-core of the primary finishes its evolution
like a single He-star and explodes as a SN, which changes the 
binary characteristics suddenly. Throughout this paper, as well as
Paper III, we assume that the neutron star formed in the SN explosion 
has a mass of 1.4 \Msun\, (as we did in Paper I)
and rest of the mass of the He-core is lost from the 
system. If this mass loss is sufficiently large (which it can be 
depending on the details of the primordial binary, despite the heavy 
mass loss from the primary during the previous CE phase), the binary 
is disrupted. If the SN explosion is symmetric, a mass loss of more 
than half of the total mass of the system will unbind the 
binary. For smaller mass losses, the parameters of the post-SN binary 
are related in a simple way to those of the pre-SN binary. However,
natal kicks given to the neutron star due to asymmetric SN explosions
alter this simple picture considerably. A quantitative general 
treatment of the binary-parameter changes including all of the above
points has been given in Sec.3.2 of Paper I, details of the computation
(particularly those of the averaging process over the distribution of
natal kicks) being given in Appendix A of that paper.   

We continue here with the notation introduced in the above places in 
Paper I. In particular, in averaging over the 3D isotropic Maxwellian 
distribution $\propto v^2\exp(-v^2/2\sigma^2)$ of the natal kick 
velocities $v$, we introduced an upper limit $v_{up}$ where we truncated 
the processes of averaging over this kick-velocity distribution (see 
Eq.(A2) of Paper I): this limit corresponds to the point at which the 
post-SN system becomes just unbound (see Eq.(A1) of Paper I), 
so that extending the integration above this limit would incorrectly 
include the unbound systems also, which in reality must be excluded.
Further, for algebraic convenience, we expressed this upper limit in
the form $v_{up}\equiv(\sqrt{2}\sigma)f$, and worked with the
parameter $f$. As shown in Appendix A of Paper I, the 
distribution-averaged kick-velocity square $v_k^2$ could then be 
expressed in the simple form $v_k^2 = \sigma^2h(f)$, where $h(f)$
was a function of $f$ alone. As detailed there and in Fig.13 of 
Paper I, two regimes of behavior were clearly shown by $h(f)$: for 
$f < f_c$, $h(f)\propto f^2$, and for $f > f_c$, $h(f)\approx$ 
constant, with a critical value $f_c\approx\sqrt{2.5}$. The 
transition region between the two regimes is very narrow. This clear
separation of regimes made the calculation of $v_k^2$ easy.
     
Here in the case of pre-LMXBs, $v_{up}$ has a relatively large range 
$\sim 50-400$ km/s. But the values $\sigma = 26.5$ km/s and 
$\sigma = 265$ km/s, corresponding to ECSN and ICCSN respectively, 
still give values of $f$ which are on opposite sides of $f_c$. 
Due to the extremely narrow transition region in $f$, these two cases 
can, therefore, still be treated in a way similar to that used in the 
treatment used for HMXBs in Paper I. Detailed calculations show that 
$\sigma = 26.5$ km/s gives results extremely close to no-kick case. 
Therefore, for our LMXB work here we need only calculate the 
large-kick ICCSN case, \ie, $\sigma = 265$ km/s, explicitly in
addition to the no-kick case, since the latter serves as an excellent 
description of the small-kick ECSN case. The Jacobian $J_{SN}$
calculated in Appendix A.2 of Paper I can be used directly in this 
work for appropriate values of the relevant parameters.

\subsection{Tidal evolution}
\label{tidal}

The post-SN binary of the neutron star and its low-mass companion
is typically very eccentric, as we show in sec. \ref{prelpdf}. Such 
a system evolves due to tidal interaction on timescales of $\sim 1$ 
Myr or less. Tidal interaction has three effects on the binary: 
(a) circularization, (b) synchronisation and (c) change in the 
semimajor axis. Tidal evolution of binaries has been much 
studied in the literature, \citep{zahn, hut}, particularly in the 
context of LMXBs \citep{kalogera96a}. We treat tidal effects in 
pre-LMXBs in this work with the aid of the prescription given by 
\citeauthor{hut}, which can be used in case of large eccentricities. 
Due to the extreme compactness of the neutron star, it does not take part 
in the tidal evolution. Further, since the orientation of the orbit 
is inconsequential for our purposes here, we consider only the changes 
in the eccentricity and semimajor axis of the orbit and those in 
the rotation of the companion. Finally, we note that we are  
interested here not in the detailed evolution of these parameters, 
but rather in the relations between their initial and final values.
Such relations can be obtained by noting that the tidal evolution
preserves the total angular momentum of the system. We note here that
\citet{kalogera96a} also treated tidal evolution similarly.
However, we also try here to incorporate the effect of the companion's 
spin through the exchange of angular momenta between the orbital 
motion and the spin of the companion. Thus the effect of the tidal 
evolution can be schematically written as $J_{orb}+J_{s,spin} =$ constant. 
Using standard expressions for orbital and spin angular momenta, 
this relation can be re-expressed as
\begin{equation}
F\left[\alpha^{-1/3}(1-e^2)^{1/2}-1\right] = n_f^{4/3}(1-\alpha k).
\label{eqn:tidal}
\end{equation}

Here $\alpha\equiv n_i/n_f$, $k\equiv\Omega_i/n_i$, and $n_i$ and $n_f$ are
the average angular velocities in the initial and final states, given
by $n = \sqrt{G(M_{NS}+M_s)/a^3}$, with the masses and semimajor axes 
appropriate for the initial and final states respectively. Since the
masses do not change during tidal evolution, $n_i$ and $n_f$ differ
only through the values of semimajor axes, so that their relation
gives that between the initial and final semimajor axes, provided 
that other quantities are specified. $\Omega_i$
is the initial spin angular velocity of the companion. It is assumed
that $\Omega_f = n_f$ and $e_f=0$, \ie, the binary is circular and 
the companion is rotating synchronously at the end of the tidal 
evolution. $\Omega_i$ and therefore $k$ can be treated as a free
parameter in our calculations, with an allowed range $0 < k <
k_{max}$, where $k_{max}$ represents a maximally spinning companion 
just after the SN. Numerical calculations show that value of $k$ 
does not change $\alpha$ by a large amount. The most likely value of 
$k$ is $k_{CE}=\sqrt{1-e^2}/(1-e)^2$. Since the CE process applies a 
large frictional drag on the binary, we can safely assume that
the companion's spin period equals the orbital period at the end of the CE 
phase. Thus $n_{CE} = \Omega_{CE} = \Omega_i$ where $\Omega_{CE}$ and $n_{CE}$ 
are the spin and the orbital angular velocities at the end of the CE phase
respectively. $F$ in Eq.(\ref{eqn:tidal}) is a function of the two masses 
and hence is constant for a given system. It is given by
\begin{equation}
F = \frac{G^{2/3}M_{NS}M_s}{(M_{NS}+M_s)^{1/3}I_s},
\label{eqn:ftid}
\end{equation}
where $I_s$ is the moment of inertia of the secondary. It can be
written in terms of solar units as $I_s = i_s\Msun\Rsun^2$. 
The scaling of $i_s$ with the stellar mass is given by \citet{rucinski}.
In this work, we use an analytical fit to their data which is given
by $i_s = 0.0832M_s^{2.27}$.

Solving Eq.(\ref{eqn:tidal}) for $\alpha$ gives the relation between
orbital separations before and after the tidal evolution. The
companion mass $M_s$ is of course unchanged during the process, and 
the \emph{initial eccentricity} is taken as the third dummy parameter
for calculational ease. Thus the transformation is essentially only in
one parameter, \ie, the orbital separation, described by $\alpha$.
The Jacobian for the inverse transformation is then given by
\begin{equation}
J_{tid} = \sqrt{\frac{a_{psn}}{a(1-e^2)}}\left[1-\frac{3}{f}n_f^{4/3}\right].
\label{eqn:jtid}
\end{equation}

\section{Constraints on Pre-LMXB Parameters}
\label{primoconstr}

Formations of LMXBs are very low-probability events. Various system 
parameters need to have values in very narrow ranges for the binary to
survive through the various phases of the evolutionary process. 
Constraints on the allowed ranges of these binary parameters at 
different stages of evolution can be translated to those on primordial 
binary parameters, using the transformations described in 
Sec.\ref{paramchange}. Such constraints were first discussed by 
\citet{kalogera98}. Let us first study the constraints which are 
relevant at pre-LMXB stage, identify the allowed zones, and calculate 
the probability of formation. These constraints can be expressed as follows:

\begin{enumerate}
\item \emph{The primary must fill its Roche lobe}. This ensures
the formation of a CE, which is an essential step towards pre-LMXB
formation. In this work, we adopt a very conservative limit for this
by setting $a_0r_L(1/q) = R_{BAGB}$, where $R_{BAGB}$ is the radius
of the primary at the base of the asymptotic giant branch. 
Analytical fits given
by \citet{hurleysse} are used to calculate $R_{BAGB}$. This gives
an upper limit on $a_0$ as a function of the metallicity of the
primary for given primary and secondary masses.
Fig. \ref{fig:alim}(a) shows this upper limit as a function of
the mass of the secondary, for $M_p=12\Msun$. A representative
lower limit for $\alpha\lambda=1$ is also shown. It can be easily
seen that higher metallicity allows higher phase-space area, thus
producing a larger number of systems, although this condition is
not very restrictive.

\insertdoblfig{scale=0.28, angle=270}{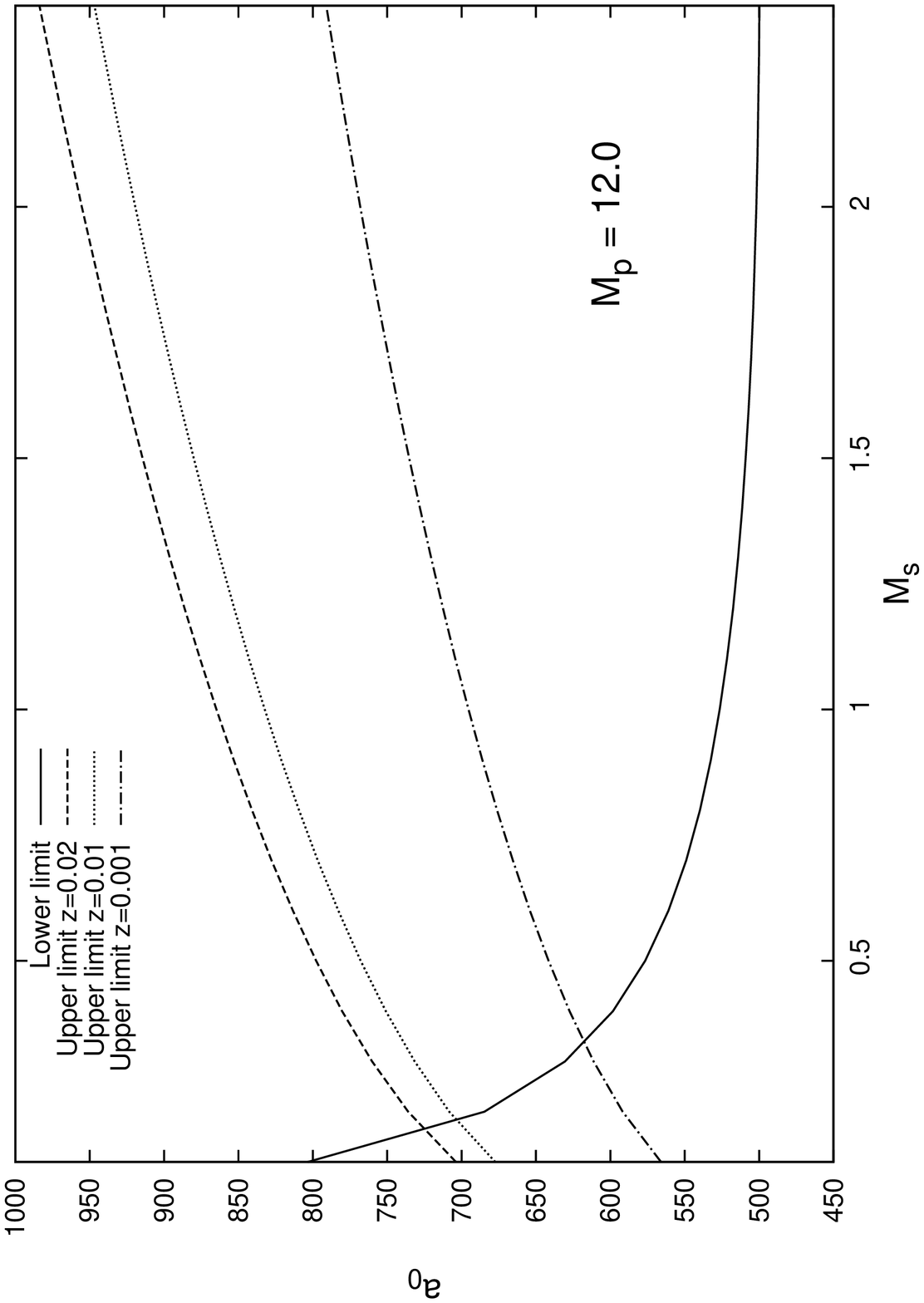}
              {scale=0.28, angle=270}{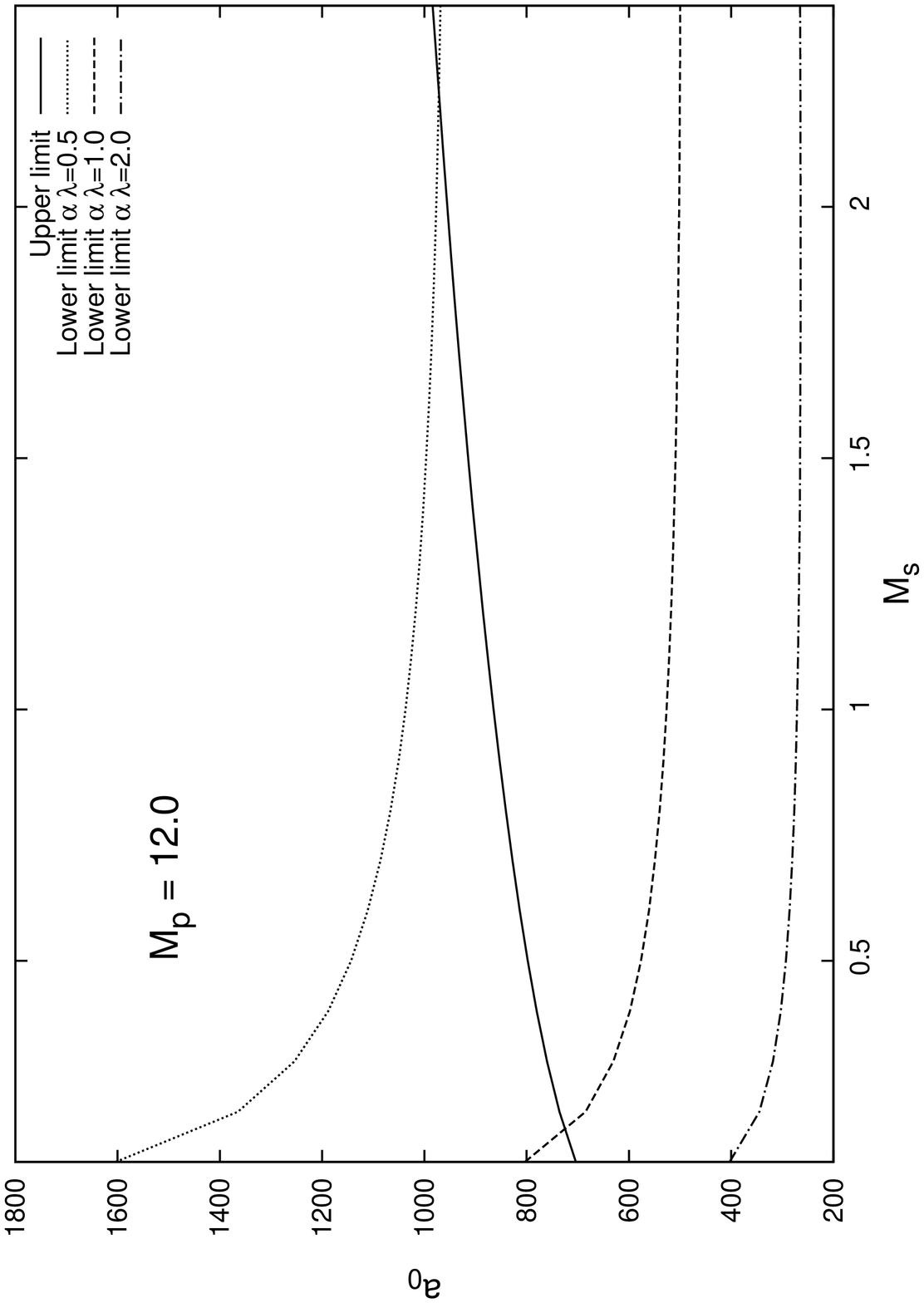}
              {\it Constraints on $a_0$ as a function of $M_s$.
                (a) left panel: upper limit for different z values.
                (b) right panel: lower limit for different CE-parameter
                values. Both plots for $M_p=12\Msun$.}
              {fig:alim}

\item \emph{The post-CE system must be detached}. This ensures
uninterrupted evolution of the He-core of the primary, resulting
in the formation of a neutron star. This is a twofold constraint,
requiring that the radii of both the secondary and the He-core must be
less than their respective Roche-lobe radii. Although this constrains
the post-CE orbital separation from below, we can convert it to a 
constraint on $a_0$ using eqn. \ref{eqn:cechange}. This constraint 
depends, of course, upon the value of the CE parameter, as shown 
in Fig.\ref{fig:alim}(b). Note that a representative upper limit, 
corresponding to $z=0.02$ (taken from the left panel (a) of this
figure), is repeated in this panel for comparison, and similarly a 
representative lower limit, corresponding to $\alpha\lambda=1$ 
(taken from this panel), is repeated in the left panel (a) of this 
figure for the same purpose. 

It is clear from the above that very inefficient
CE processes prohibit most of the parameter space, since the lower
limit is larger than the the upper limit over a large range of $M_s$.
For values of the CE parameter above unity, however, a large range 
is allowed and the constraints are not very restrictive. 
We also note that changes in the CE parameter make larger changes 
in the allowed range of $a_0$, making it the more dominant parameter 
in determining the PDF.

\item \emph{The binary must survive the SN explosion}. Disruption
of the binary due to sudden mass loss is discussed above in 
Sec.\ref{sn}, with reference to detailed calculations in Paper I.
This requirement puts a lower limit on the allowed values of $M_s$, 
or equivalently on those of $q$, as functions of $M_p$. We set an 
absolute lower limit on $M_s$ as 0.1\Msun. Thus the limit obtained
from the condition of SN survival is applicable only above a certain
value of $M_p$, below which the entire range is allowed. We set the 
upper limit on $M_s$ as 2.5\Msun\, matching with the constraints used by
\citet{grimm01}. Fig.\ref{fig:mslim} depicts these constraints.
It can be seen that, for neutron-star LMXBs, the progenitor primary 
must be less massive than $\approx 21\Msun$. 

\insertfig{scale=0.5, angle=270}{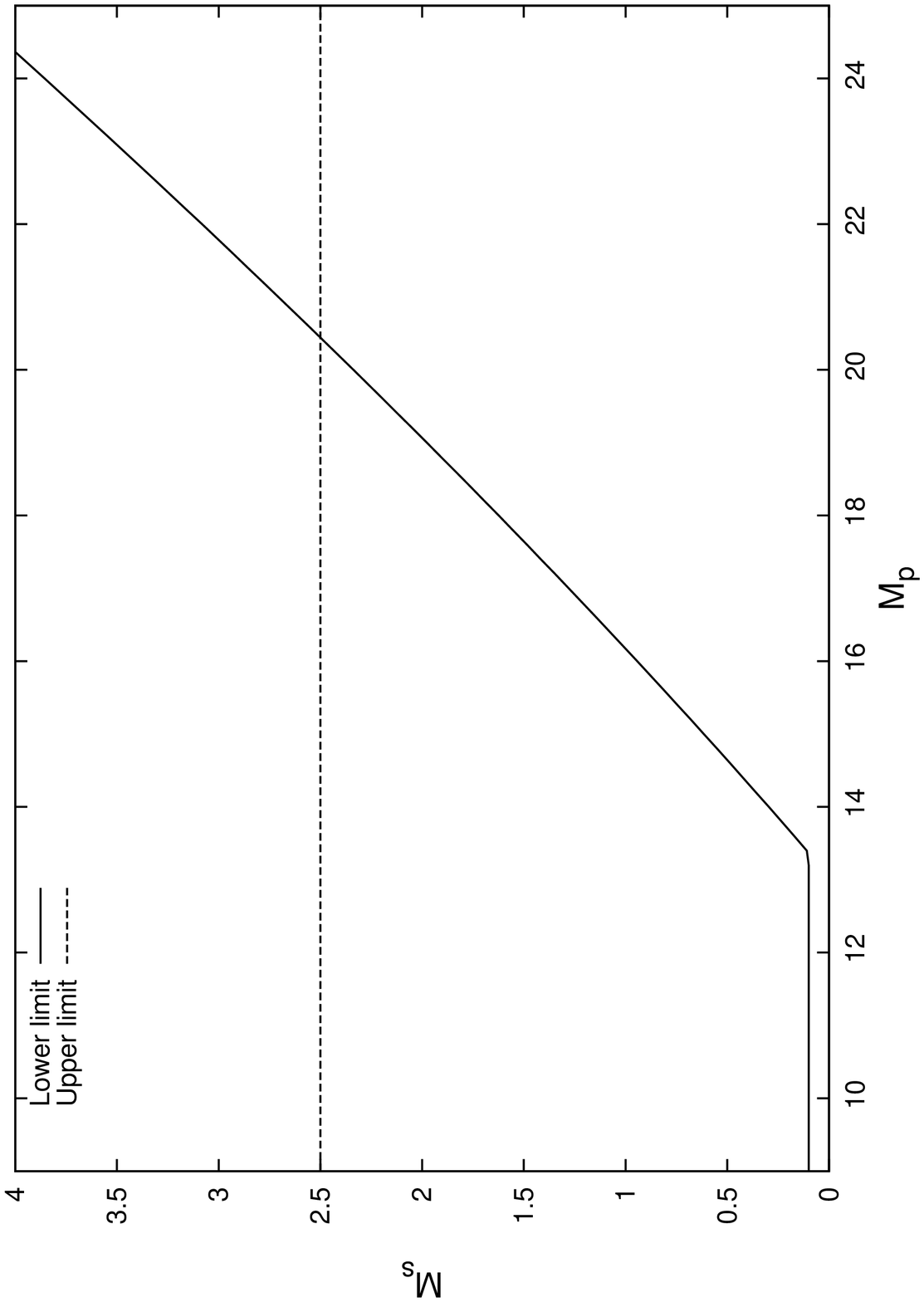}{\it Constraints on the
  allowed range of $M_s$ as a function of $M_p$.}{fig:mslim}

\end{enumerate}

\section{Pre-LMXB Parameter Distribution}
\label{distr}

\subsection{Distribution of primordial binary parameters}
\label{primopdf}

Primordial binary distributions for LMXB progenitors are characterized 
by three parameters, \viz, the primary mass $M_p$, the 
secondary-to-primary mass ratio $q$, and the orbital separation 
($a_0$), as for the HMXB progenitors considered in Paper I.
These three parameters essentially follow the same distribution as
in case of HMXBs, \ie, an IMF describing the distribution of $M_p$, a
power-law (flat or falling) distribution for $q$ and \"Opik's law for 
$a_0$. But the allowed ranges for some of these parameters are
drastically different from those that apply to HMXB
progenitors. Especially, we note 
that the $q$-ranges for the two cases are completely disjoint, as
expected from the thermal-timescale arguments. Also, the range of
$a_0$ is more tightly constrained in the case of LMXBs due to stricter 
conditions of survival through various phases, as described in 
Sec.\ref{primoconstr}.

The requirement of a viable formation of CE constrains $q$ from 
above as $q<0.3$, which is an absolute upper limit. 
This allows us to have binaries with
companion masses such that the some of the final products may be more
appropriately called \emph{Intermediate Mass X-ray Binaries} (IMXBs),
in addition to LMXBs. The upper limit for the companion mass is 
generally taken to be in the range $1-3\Msun$ if the binary is to be
classified as an LMXB. We take this upper limit as 2.5 \Msun\ following
\citet{grimm01}. This puts another upper limit on $q$ given by $2.5/M_p$.
We find that, for the entire range of $M_p$, this limit is always
lower than the limit imposed by CE formation given above. However,
the distribution of $q$ is  poorly constrained within this range. It
has been suggested by \citet{sana} that the uniform $q$-distribution
observed for $q>0.2$ can be extended to the lower $q$-values. For 
LMXB/IMXB population studies, this uniform $q$-distribution has been 
indeed been used by \citet{pfahl03}. By contrast, a
$q^{-2.7}$-distribution, \ie, a rather steeply falling power-law,
was used in the comprehensive LMXB work of \citet{kalogera98}. We in this
work use the general form $\propto q^{\beta}$ and take $\beta=0$
and $\beta=-2.7$ to accomodate both the approaches mentioned above,
as also to explore any other power-law, if necessary.

The distribution of the orbital separation is taken as \"Opik's law,
\citep{opik} as per current norm. We note however that the tight
constraints in the LMXB-formation process described in section 
\ref{primoconstr} severely limit the allowed range of $a_0$ which can 
eventually produce viable pre-LMXBs. It is not impossible that 
fluctuations within this small range may have been overlooked while 
deriving \"Opik's law for a much wider range of separations. However, 
due to the lack of a close coverage of data over small ranges of 
separations, we find it most prudent to assume at this time that 
the wide-range log-uniform distribution also applies over the smaller 
range relevant for our purposes here. At a minimum, it certainly 
gives an indication of the average trend.
  
The distribution of the primary mass is given by the IMF 
\citep{salpeter, kroupa93, kroupa03}. We in this work use the IMF
given by \citeauthor{kroupa03} since it is applicable for a 
relatively wide range of masses and can be extended to the 
lowest mass ranges relevant for the secondary in these calculations. 

The PDF of the primordial binaries can therefore be written as
\begin{equation}
f_{prim}(M_p, q, a_0) = \frac{1}{N} \frac{\mbox{IMF}(M_p)\;q^{\beta}}{a_0}
\label{eqn:primpdf}.
\end{equation}
Here, $N$ is the normalization parameter defined such that when the
above PDF is integrated over the relevant ranges of all the 
parameters, it yields unity. 

\subsection{Transformation of PDF to pre-LMXB stage}
\label{prelpdf}

The Jacobian formalism described in Paper I can be used to
transform the distribution of primordial binaries to the distribution
of pre-LMXBs. Formation of pre-LMXBs proceeds through the three stages
of parameter changes described in Sec.\ref{paramchange}. The binary at
each stage is defined by three parameters, which can be connected to
the parameters of the next stage. We start with the transformation
from primordial binary to post-CE binary. The three parameters describing
the primordial binary are $(M_p, q, a_0)$ and those describing the post-CE
binary are $(M_{p,c}, M_s, a_{CE})$. The transformation relations are 
given by Eqs. \ref{eqn:cechange}, \ref{eqn:mcore} and the relation 
between $q$ and $M_s$ is of course $M_s = qM_p$. The Jacobian for the 
inverse transformations is given by Eq.(\ref{eqn:jce}).

The next stage of parameter change occurs at the SN explosion. 
The post-SN binary is described by $(M_s, a_{psn}, e)$. We first note
that $M_s$ is unchanged during this transformation. 
The inverse transformations for
the other two parameters and the Jacobian for that are described 
in Appendix A of Paper I. As noted in Sec.\ref{sn}, we perform the
calculations for ICCSN $\sigma=265$ km/s case and for the no-kick
case, which also represents the ECSN case with $\sigma=26.5$ km/s.

For the post-tidal binary, only two parameters, \viz, $(M_s, a)$ are
sufficient to describe the system. To describe the immediate post-SN
system, however, we also need the orbital eccentricity $e$, \ie, a total of
three parameters. We do the transformation of parameters and
distributions between pre- and post-SN systems as described, and then
integrate over the eccentricity in order to arrive at the post-tidal binary 
parameters. The distribution of the immediate post-SN eccentricity is shown 
in Fig.\ref{fig:edfplm}. The small kicks characteristic of ECSN
(whose effects are essentially identical to those for no kicks, as 
explained above) give an $e$-distribution 
which peaks at low eccentricities, while  
the large kicks characteristic of ICCSN give high eccentricities with
the $e$-distribution peaking at a rather large value, as expected. 
We note that the positions of the peaks of the $e$-distributions
obtained here for both ECSN and ICCSN are similar to those 
obtained for HMXBs (see Paper I). This is particularly true in the 
latter case, although the distribution for LMXBs appears to be 
much wider than that for HMXBs. We emphasize again here that the 
immediate post-SN $e$-distribution is not amenable to observation,
though indirect clues on eccentricities of pre-LMXBs can be obtained
from observations of individual X-ray binaries if they can be
plausibly identified as pre-LMXBs, as described by \citet{harshal1E}.

\insertfig{scale=0.5, angle=270}{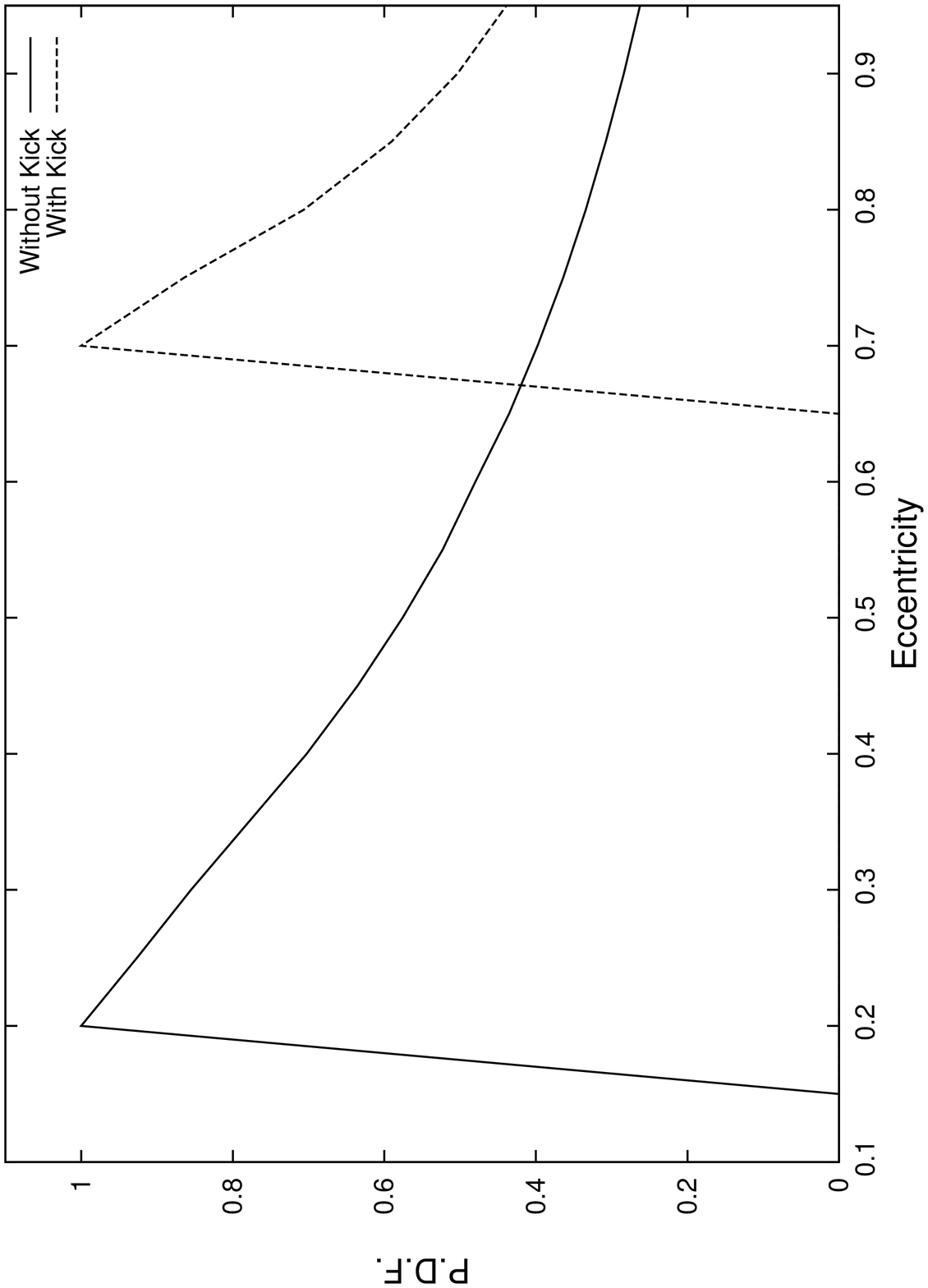}{\it Distribution of 
post-SN eccentricities. For reasons explained in the text, 
the ECSN case is labeled ``Without Kick'', and the ICCSN case, 
``With Kick''.}{fig:edfplm}

The change in the semimajor axis is obtained by solving Eq.
(\ref{eqn:tidal}) numerically for $\alpha$, which gives $a/a_{psn}$.
The Jacobian for this transformation is given by Eq.(\ref{eqn:jtid}).

The primordial-binary PDF, $f_{prim}$ given by Eq.(\ref{eqn:primpdf}) 
can now be transformed into the pre-LMXB PDF using the Jacobian 
transformation, which can be schematically written as:
\begin{equation}
f_{ptid}(M_s, a, e) = f_{prim}(M_p, q, a_0) J_{CE}J_{sn}J_{tid}.
\label{eqn:ptidpdf}
\end{equation}

The pre-LMXB PDF is finally obtained by integrating $f_{ptid}$ over 
eccentricity:
\begin{equation}
f_{plm}(M_s, a) = \int_0^1f_{ptid} \; de.
\label{eqn:plmpdf}
\end{equation}

\subsection{Properties of the pre-LMXB PDF}
\label{prelprop}

The pre-LMXB PDF is a bivariate function of $M_s$ and $a$. We display this
bivariate PDF in Figs.\ref{fig:plmfl}(a, b) and \ref{fig:plmpl}
(a, b) for the four possible cases arising out of the  
two $q$-distributions (\ie, flat or steeply falling power-law) and the 
two SN-kick situations (\ie, no-kick/small ECSN kick or large ICCSN kick)
that we explore in this work. In these 3-D displays, we have attempted to
use the optimal viewing angles for bringing out the most essential 
features of the distributuion in each case, so that this viewing angle 
is not the same in all cases. 
 
\insertdoblfig{scale=0.28, angle=270}{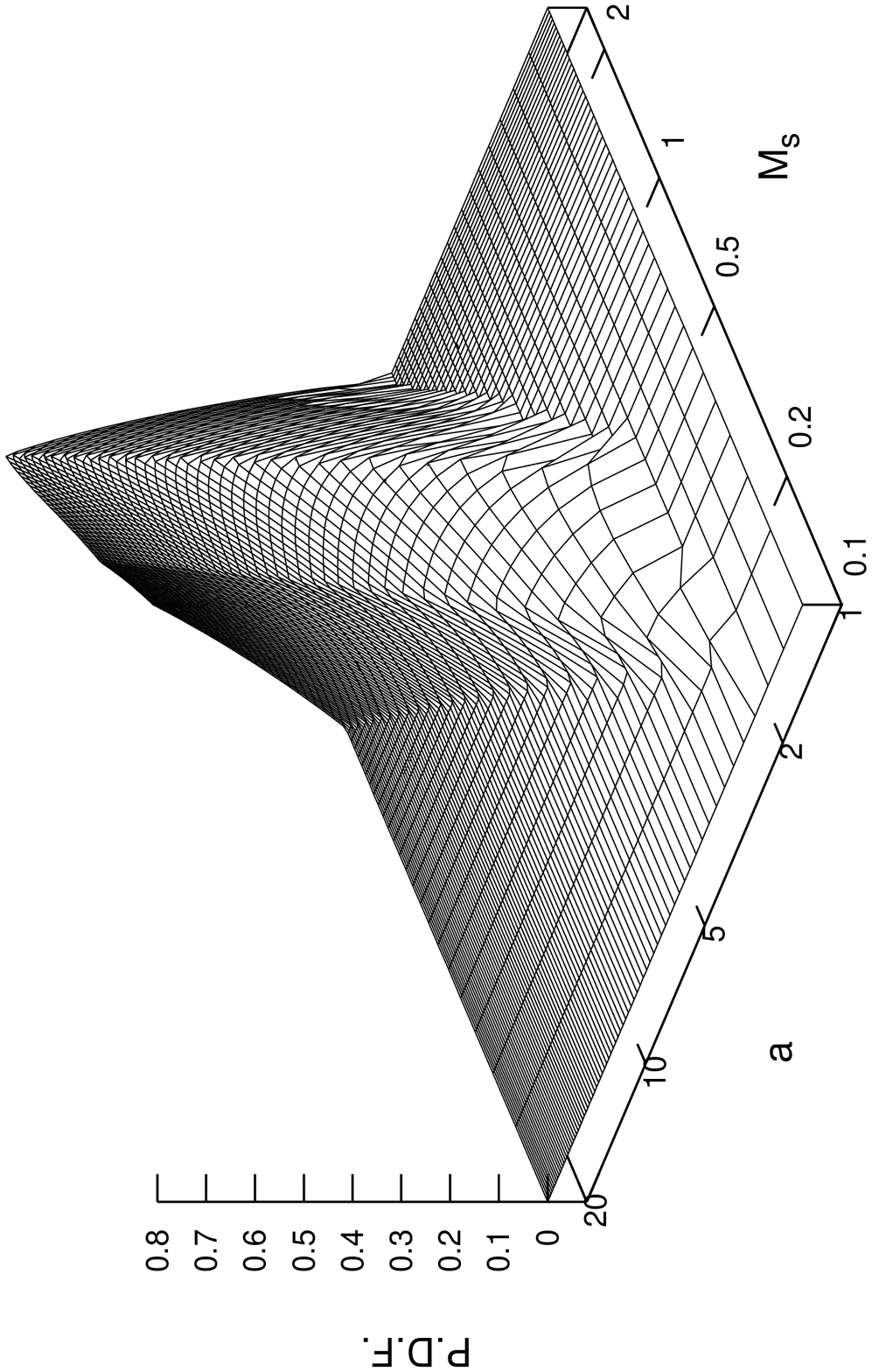}
              {scale=0.28, angle=270}{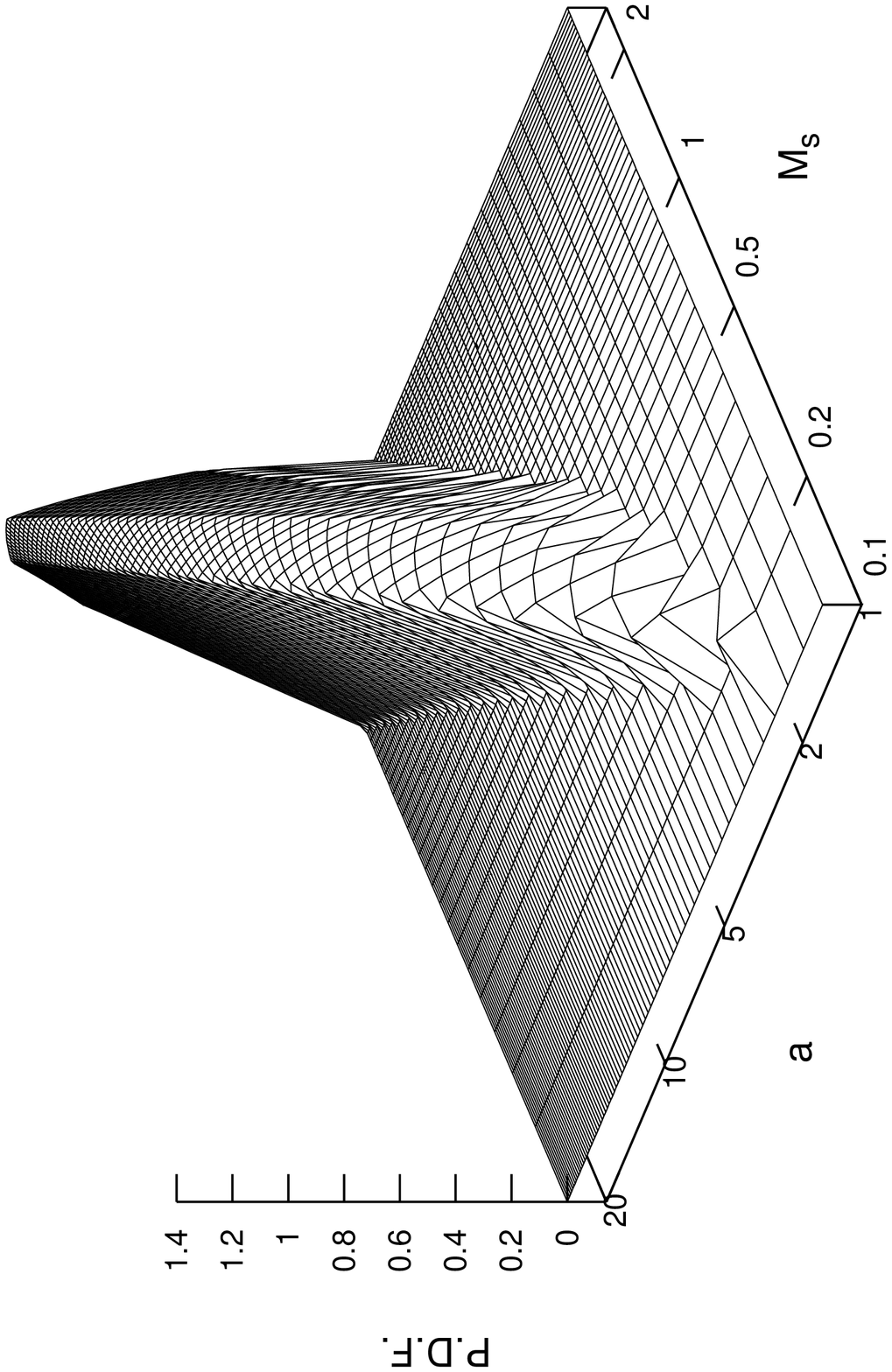}
              {\it Bivariate pre-LMXB distribution as a function of $M_s$ and
                $a$ for a flat $q$-distribution.
                Left (panel a): without kick, Right (panel b): with kick}
              {fig:plmfl}

\insertdoblfig{scale=0.28, angle=270}{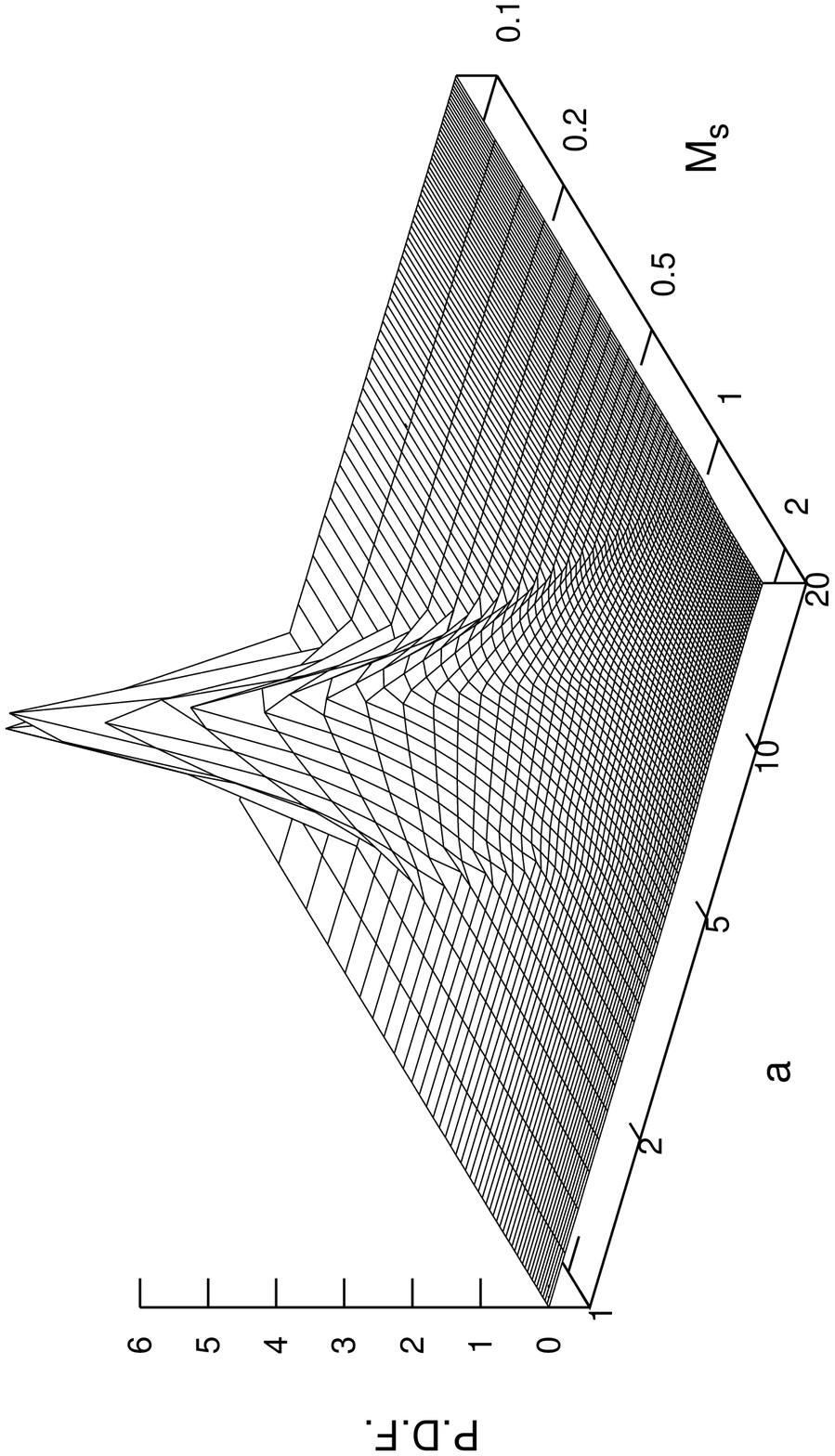}
              {scale=0.28, angle=270}{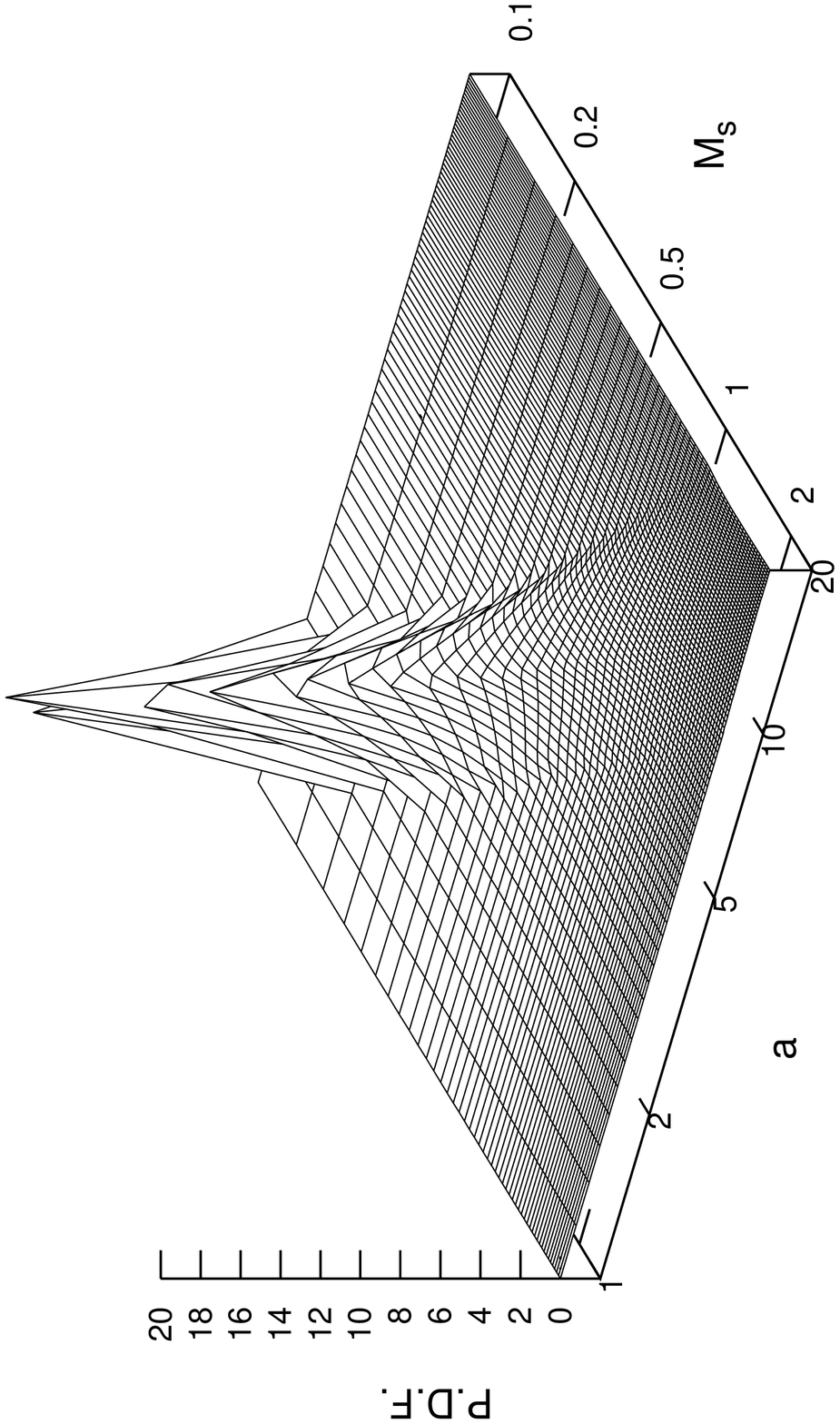}
              {\it Bivariate pre-LMXB distribution as a function of $M_s$ and
                $a$ for a falling power-law $q$-distribution.
                Left (panel a): without kick, Right (panel b): with kick}
              {fig:plmpl}

The individual (monovariate) PDF for a given variable can then be
obtained by further integration of the above bivariate PDF
over the other variable. Since $f_{prim}$ has been normalized over the
allowed range for the viable pre-LMXB formation, $f_{plm}$ is  
normalized automatically. However, any
further integration requires the knowledge of the appropriate limits for
the two parameters of the pre-LMXB systems. The range of $M_s$ is [0.1-2.5], 
which comes from various constraints described in Sec.\ref{primoconstr}.
The distribution of $a$ is obtained by integrating $f_{plm}$ over this
range of values of $M_s$. The integration range of $a$ is chosen as
[0-20] for obtaining the PDF as a function of $M_s$. Note that the
lower $a$-limit is not very restrictive, as the PDF must fall to zero at
$a \approx 1.5R_c$, so that Roche lobe contact does not occur just
after the SN.  The upper $a$-limit comes from the fact 
that wider systems will not come into Roche-lobe contact in a Hubble
time. The actual value chosen is typical,
corrsponding to the range of values of the pre-LMXB companion mass. 
We note here that $f_{plm}$ does not fall to zero at this upper limit, 
so that we are excluding at this point a few systems which are not 
expected to reach the LMXB phase. Consequently, the individual 
(monovariate) PDF is not automatically normalised, unlike earlier.

The individual PDFs thus obtained depend upon the parameters which decide
the strengths of the various steps involved in the formation of pre-LMXBs.
These parameters are (a) the CE parameter (default value 1.0) (b) metallicity
(default value 0.02) (c) exponent of the mass-ratio PDF, $\beta$, and   
(d) the scenario for the SN-kicks (\ie, no-kick/small ECSN kick or
large ICCSN kick).

\subsubsection{Companion-mass distribution}

\insertfig{scale=0.5, angle=270}{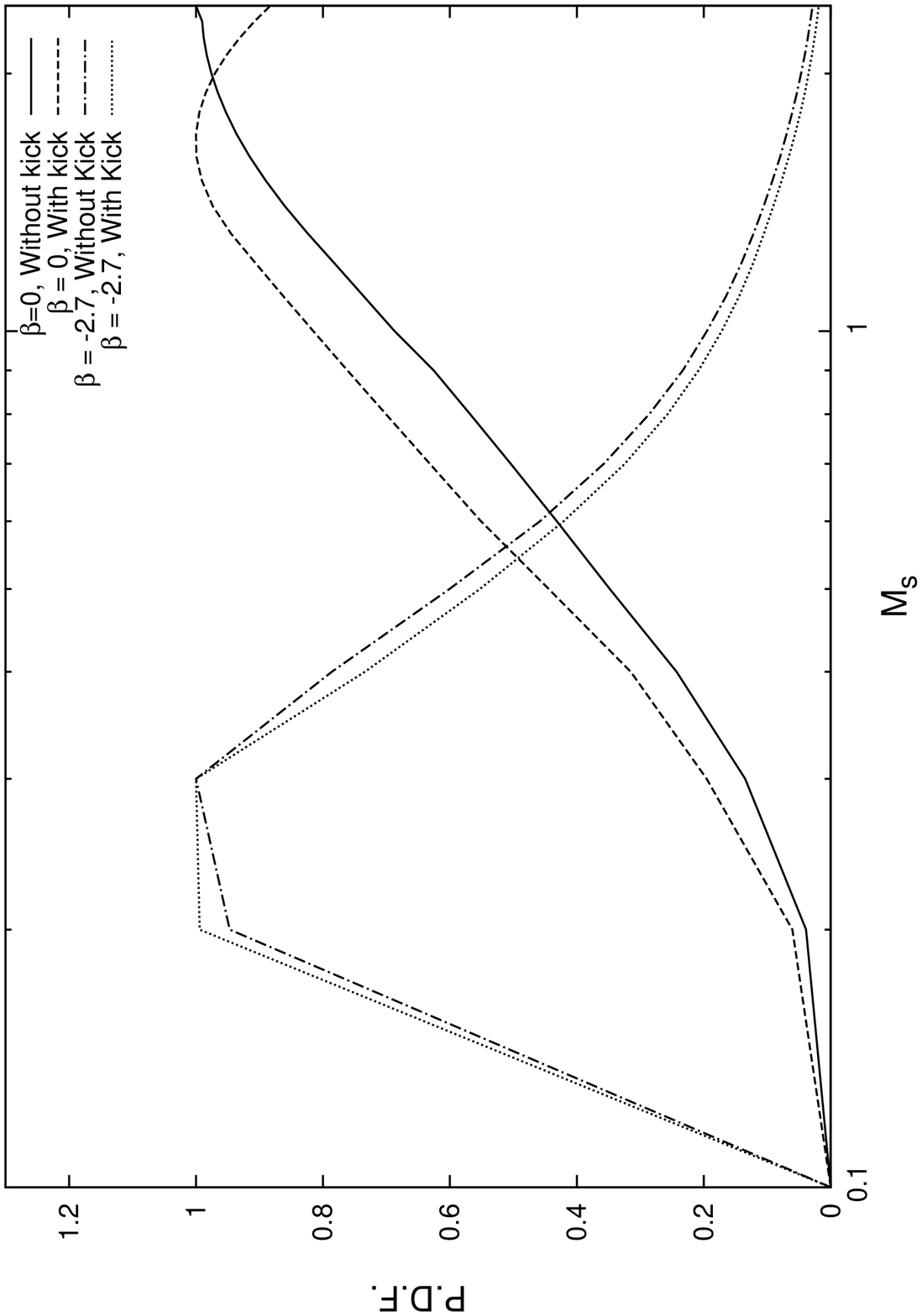}{\it Distribution of the companion
  mass for the four cases displayed in Figs.\ref{fig:plmfl}(a, b) and 
  \ref{fig:plmpl}(a, b). Cases coded by line styles as indicated.}{fig:pmsqpl}

The distribution of the companion mass is not directly affected by the
CE parameter and the metallicity. The effects of these parameters
enter indirectly through the allowed phase-space area, since these 
parameters affect the allowed range of $a$. By contrast, the
primordial mass-ratio distribution exponent $\beta$ 
affects the distribution of the companion mass directly. Fig. 
\ref{fig:pmsqpl} shows the companion-mass PDF for $\beta = 0$ (flat 
distribution) and $\beta = -2.7$ (power-law distribution), clearly
demonstrating that a falling power-law distribution makes the $M_s$ 
distribution peak at much lower values of the companion mass and 
so makes the rise to this peak faster. On the other hand, 
the shape of the PDF changes little between the different kick-scenarios, 
particularly in the $\beta = -2.7$ case. The shapes of these PDFs are
in agreement with those given in earlier works in the subject, \ie, 
\citet{pfahl03} for the $\beta = 0$ case, and \citet{kalogera98}
for the $\beta = -2.7$ case. In the former case, the agreement
is particularly striking. In the latter case, the authors gave their 
PDFs as bivariate 3D-plots, so that it is easier to compare our 
corresponding plots displayed in Figs.\ref{fig:plmpl}(a, b). 
The nearly-flat-top region in our $M_s$-distribution for $\beta =
-2.7$ and large ICCSN kicks shows up as a ``ridge'' in our bivariate 
distribution of Fig.\ref{fig:plmpl}(b), which can be compared with the 
corresponding feature in Fig.6 of \citet{kalogera98}.
                                                       
\insertdoblfig{scale=0.28, angle=270}{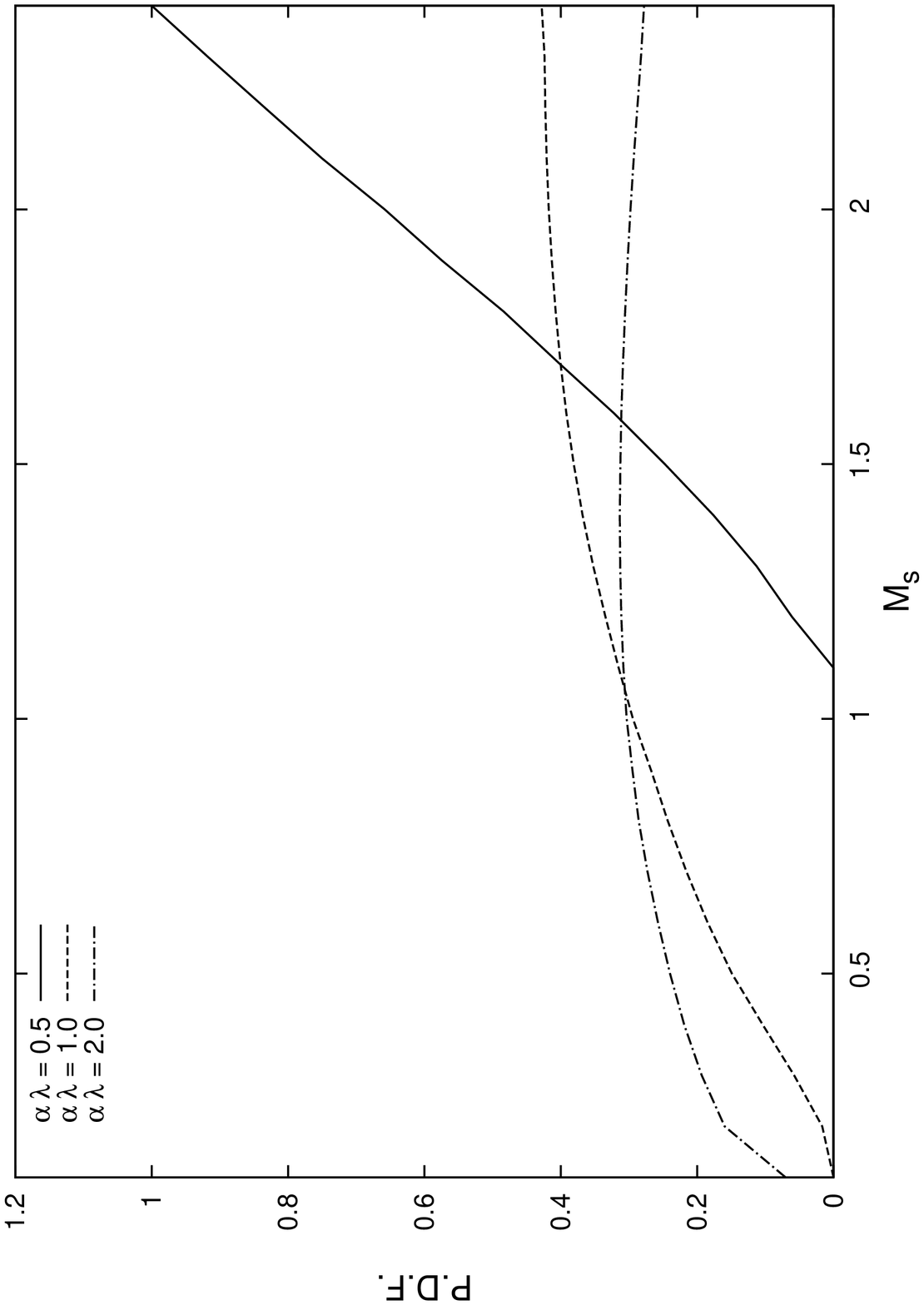}
              {scale=0.28, angle=270}{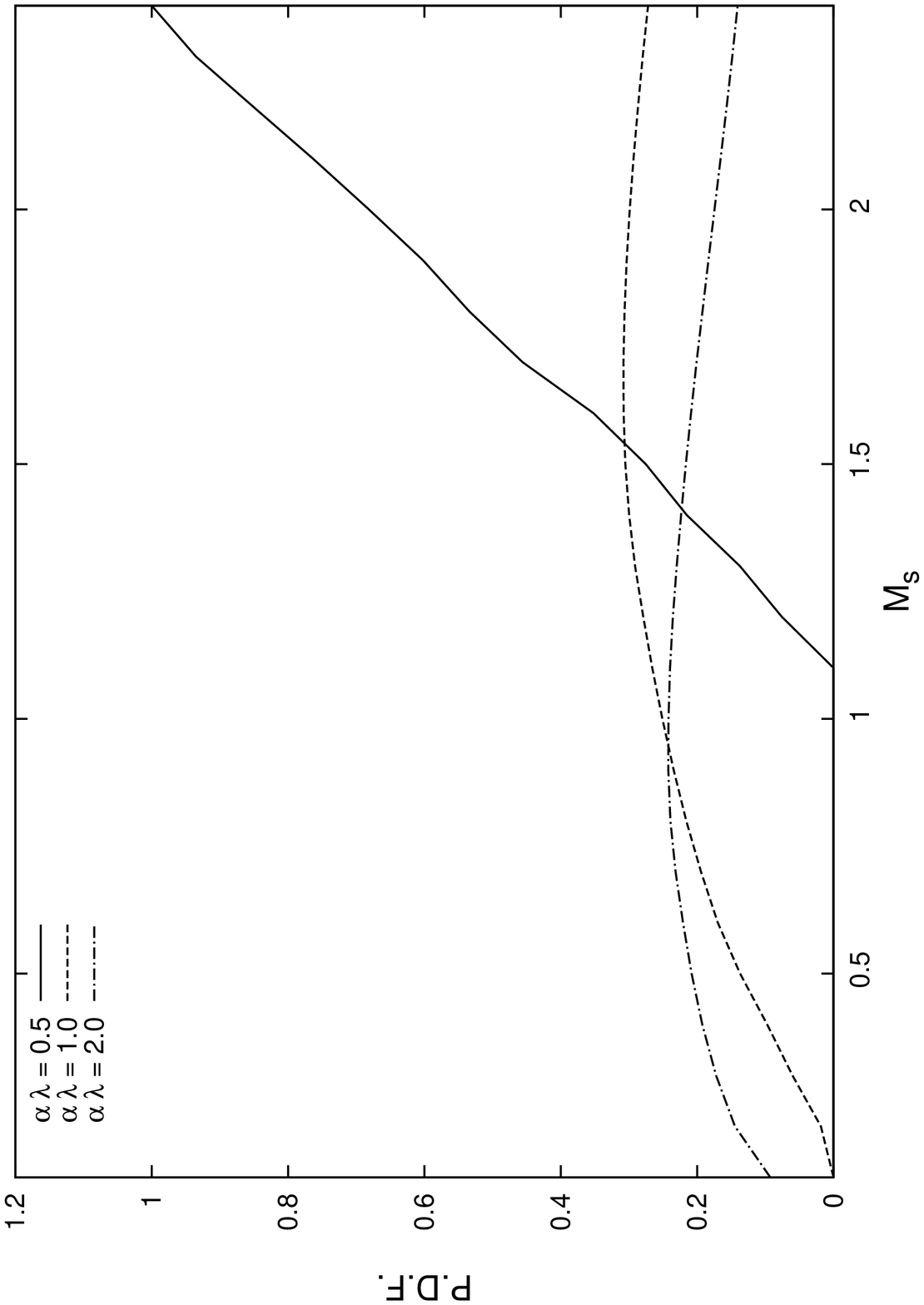}
              {\it Distribution of the companion mass for different values of
                the CE parameter $\alpha\lambda$, whose values are 
                coded by line styles as indicated. Left (Panel a):
                no-kick/small ECSN kick. Right (Panel b): 
                large ICCSN kick.}{fig:pmsce}
                                                                         
Fig.\ref{fig:pmsce} shows the variation of the companion mass PDF for 
different values of the CE parameter. Drastic difference between $\alpha\lambda
=0.5 \,\rm{and}\, 2$ case can be seen. Whereas for $\alpha\lambda = 0.5$ the 
PDF is linearly rising, for $\alpha\lambda = 2.0$ it shows an early rising part 
which quickly saturates. The $\alpha\lambda =1$ case can be seen as an 
intermediate one. Again, we see that the inclusion of large kicks does
not change the shape of the PDF drastically.

\insertdoblfig{scale=0.28, angle=270}{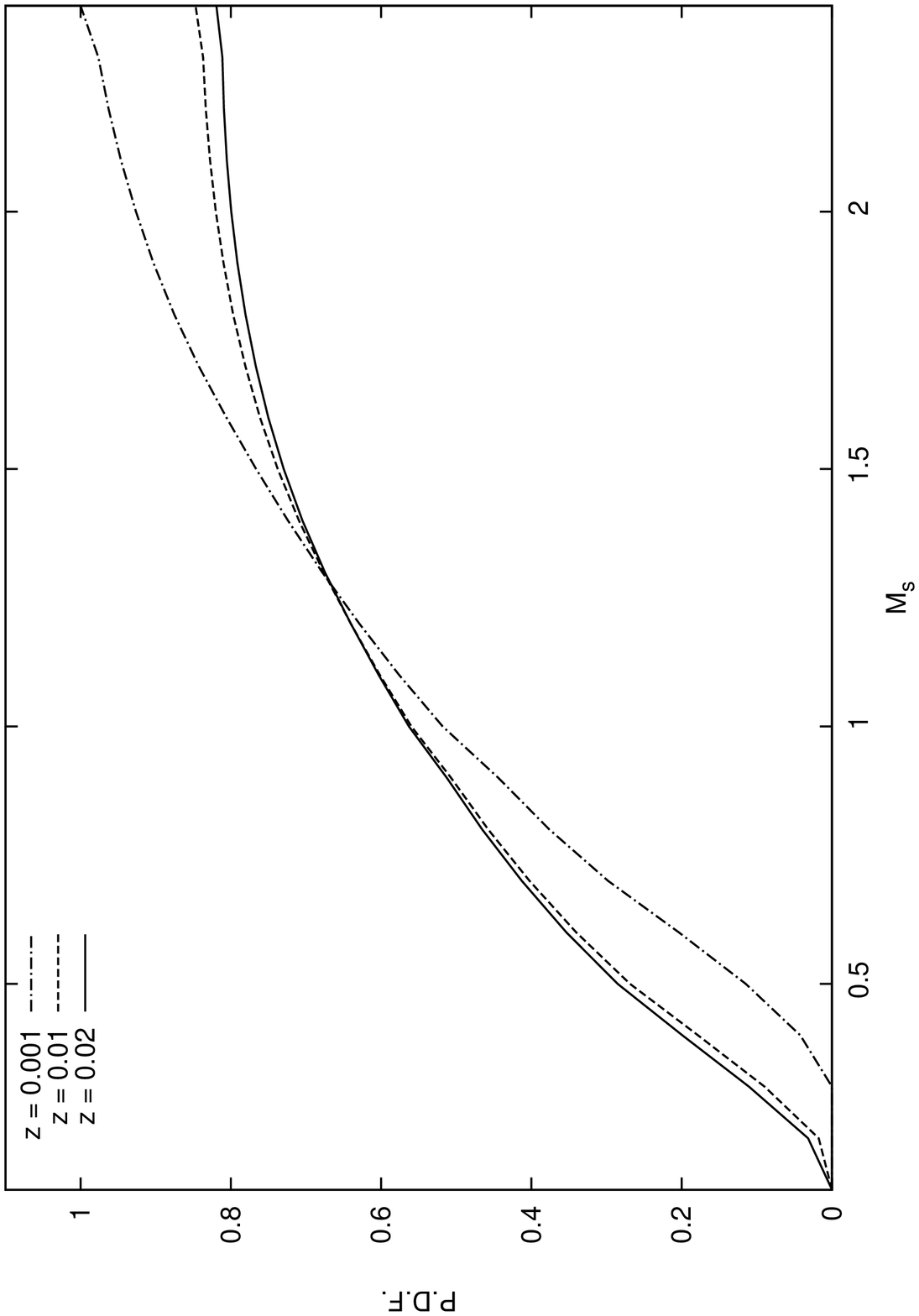}
              {scale=0.28, angle=270}{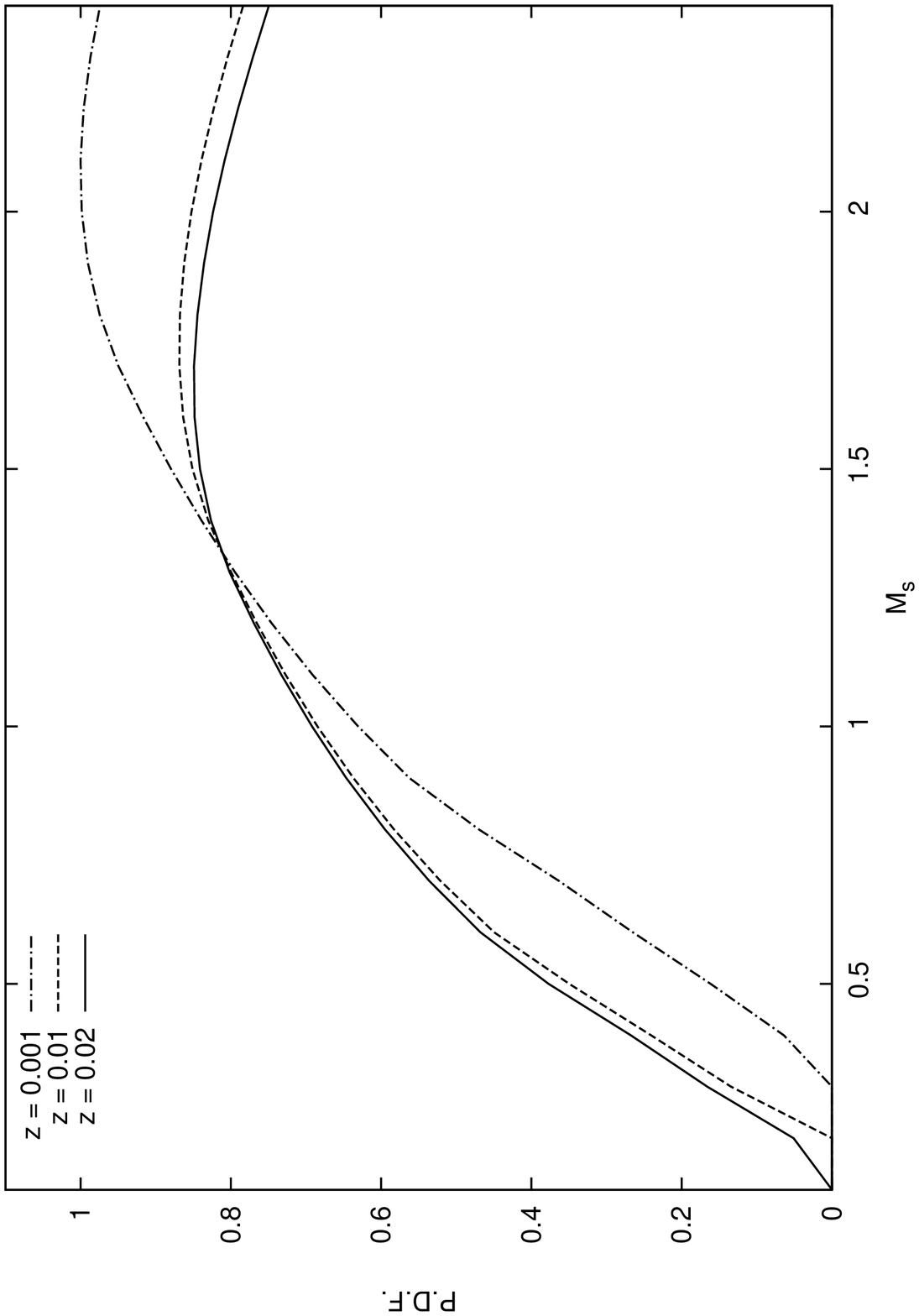}
              {\it Distribution of the companion mass for different values of
                metallicity $z$, whose values are 
                coded by line styles as indicated. Left (Panel a):
                no-kick/small ECSN kick. Right (Panel b): 
                large ICCSN kick. }{fig:pmsz}

Variations in the metallicity affect the $M_s$-PDF more weakly than 
those in the CE-parameter. This is as expected, since the effects of
both of these parameters are only through their influences on the 
allowed zones as explained above, and since the influences of the 
metallicity are much weaker. We further note that the $z=0.01$ and 
$z=0.02$ cases are not very different from each other.

\subsubsection{Distribution of orbital separation}

The distribution of the orbital separation $a$ is also affected by the above
parameters. Fig.\ref{fig:pace}(a, b) shows the variation of the 
$a$-distribution with the CE parameter for the two SN-kick scenarios 
(note that this and the next two figures show the PDF plotted on a 
logarithmic $a$-scale, \ie, $PDF = dP/d\ln a$ is displayed as a
function of $a$ plotted on a logarithmic scale). The peak of the
distribution shifts to smaller $a$-values as the CE parameter
increases, reflecting an increase in the allowed zone. The $a$-distribution
is clearly not log-uniform for either kick-scenario.

\insertdoblfig{scale=0.28, angle=270}{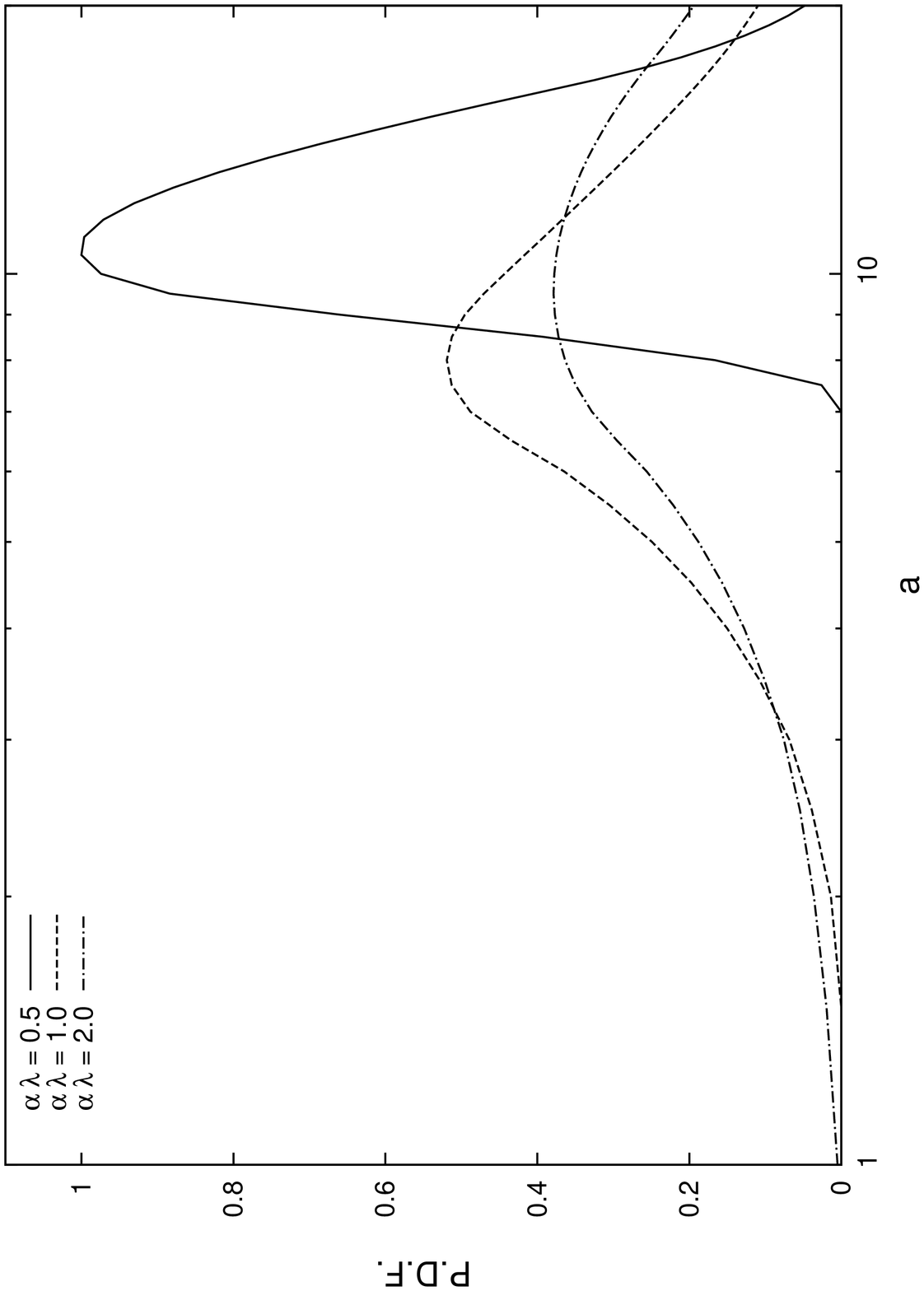}
              {scale=0.28, angle=270}{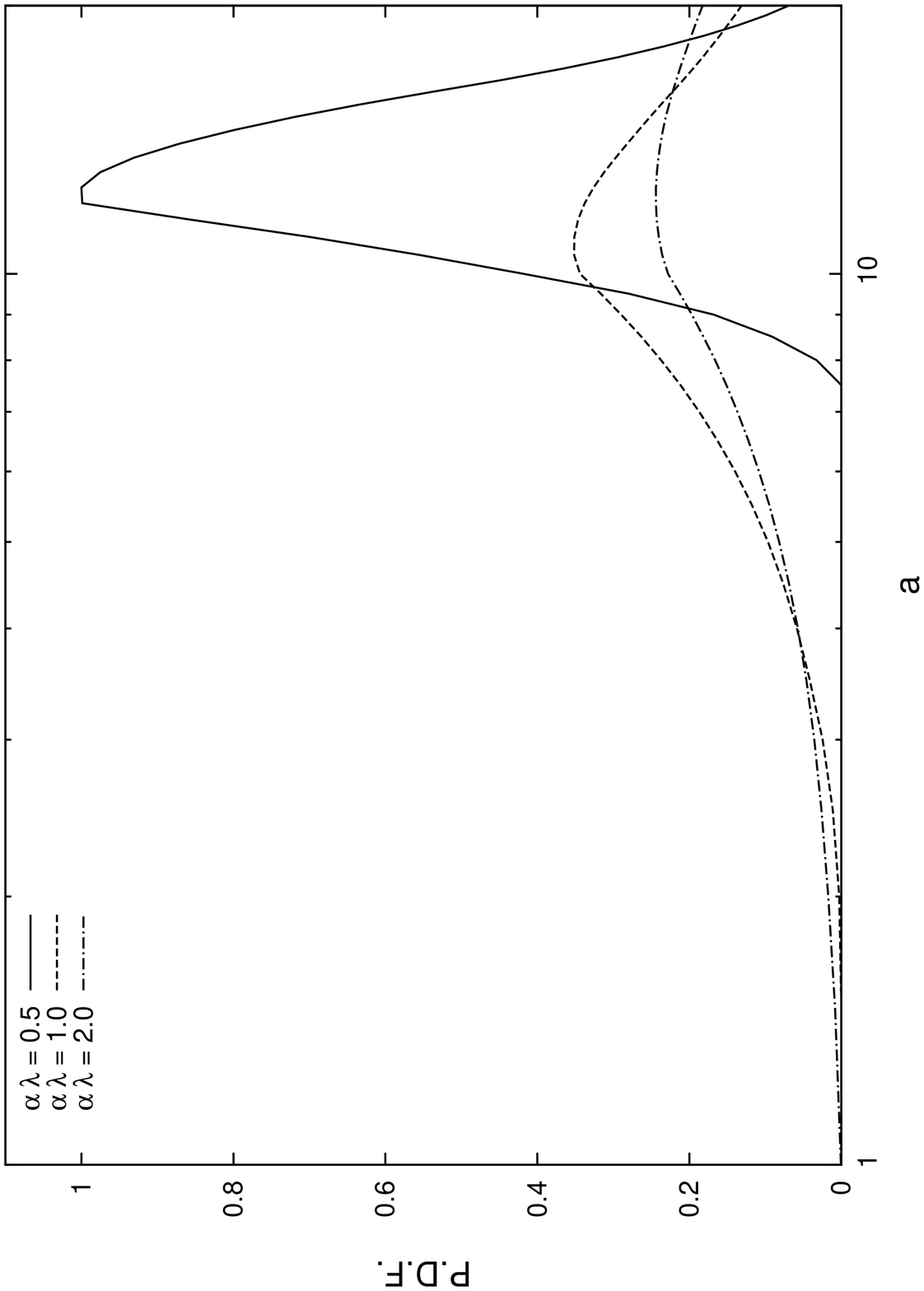}
              {\it Distribution of the orbital separation $a$ for different values 
                of the CE parameter, these values being
                coded by line styles as indicated. Left (Panel a):
                no-kick/small ECSN kick. Right (Panel b): 
                large ICCSN kick. }{fig:pace}

\insertdoblfig{scale=0.28, angle=270}{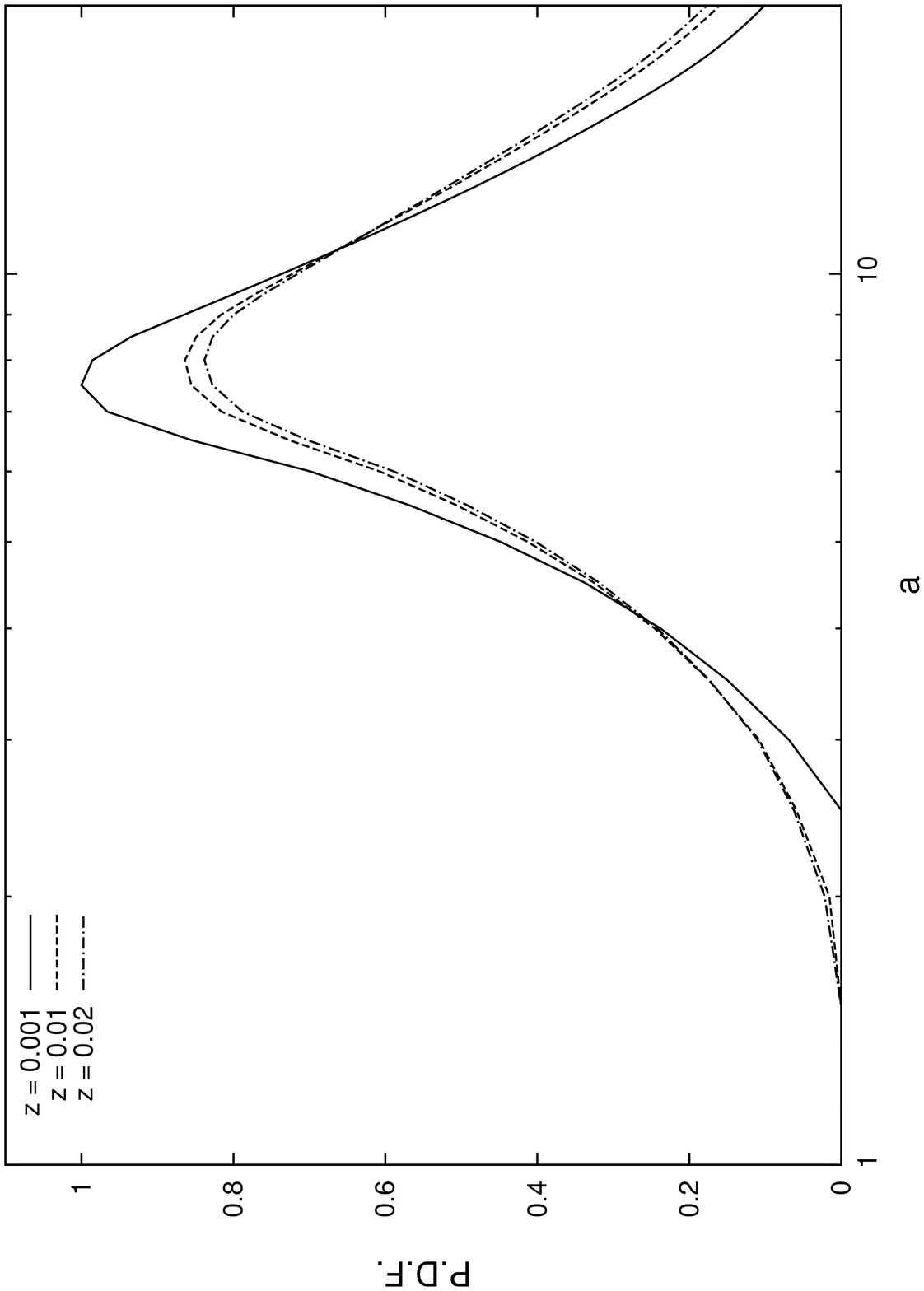}
              {scale=0.28, angle=270}{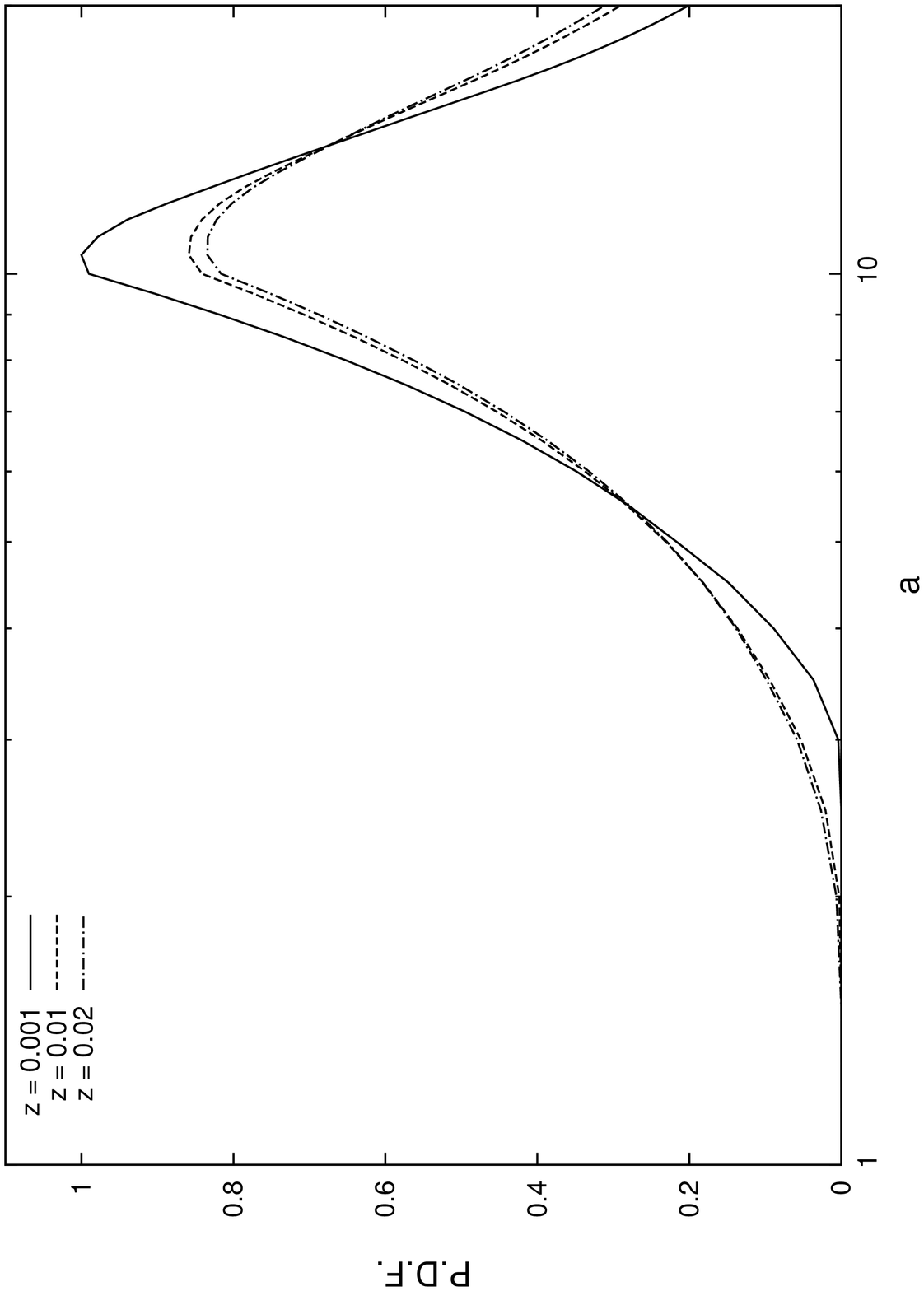}
              {\it Distribution of the orbital separation for different values 
                of metallicity, these values being
                coded by line styles as indicated. Left (Panel a):
                no-kick/small ECSN kick. Right (Panel b): 
                large ICCSN kick. }{fig:paz}

Fig.\ref{fig:paz}(a, b) shows the variation of $a$-distribution with metallicity.
Little change is seen for different values of the metallicity, with the
peak shifting slightly to the left at lower metallicities. Again, we see little
change in the shape of the PDF after the inclusion of SN-kicks.
 
Fig. \ref{fig:paqpl} shows the effect of varying $\beta$, \ie, the 
exponent of the $q$-distribution, on the $a$-distribution, which
is quite strong. The steeply-falling power law with $\beta = -2.7$ 
shifts the distribution peak to much smaller values of $a$ compared 
to the situation for the flat ($\beta=0$) $q$-distribution.
Again, we find here close similarities to previous results in the 
literature, \ie, Pfahl et al. (2003) for the $\beta = 0$ case, and 
Kalogera \& Webbink (1998) for the $\beta = -2.7$ case, and again
the former similarity is particularly striking. We emphasize that
the general rise-and-fall shape of the $a$-distribution for pre-LMXBs
seems both quite generic and confirmed by all previous calculations 
known to us, and that this shape stands in contrast to the generically 
flat or nearly-flat shape at intermediate $a$'s that we found for 
pre-HMXBs and HMXBs in Paper I. 

\insertfig{scale=0.5, angle=270}{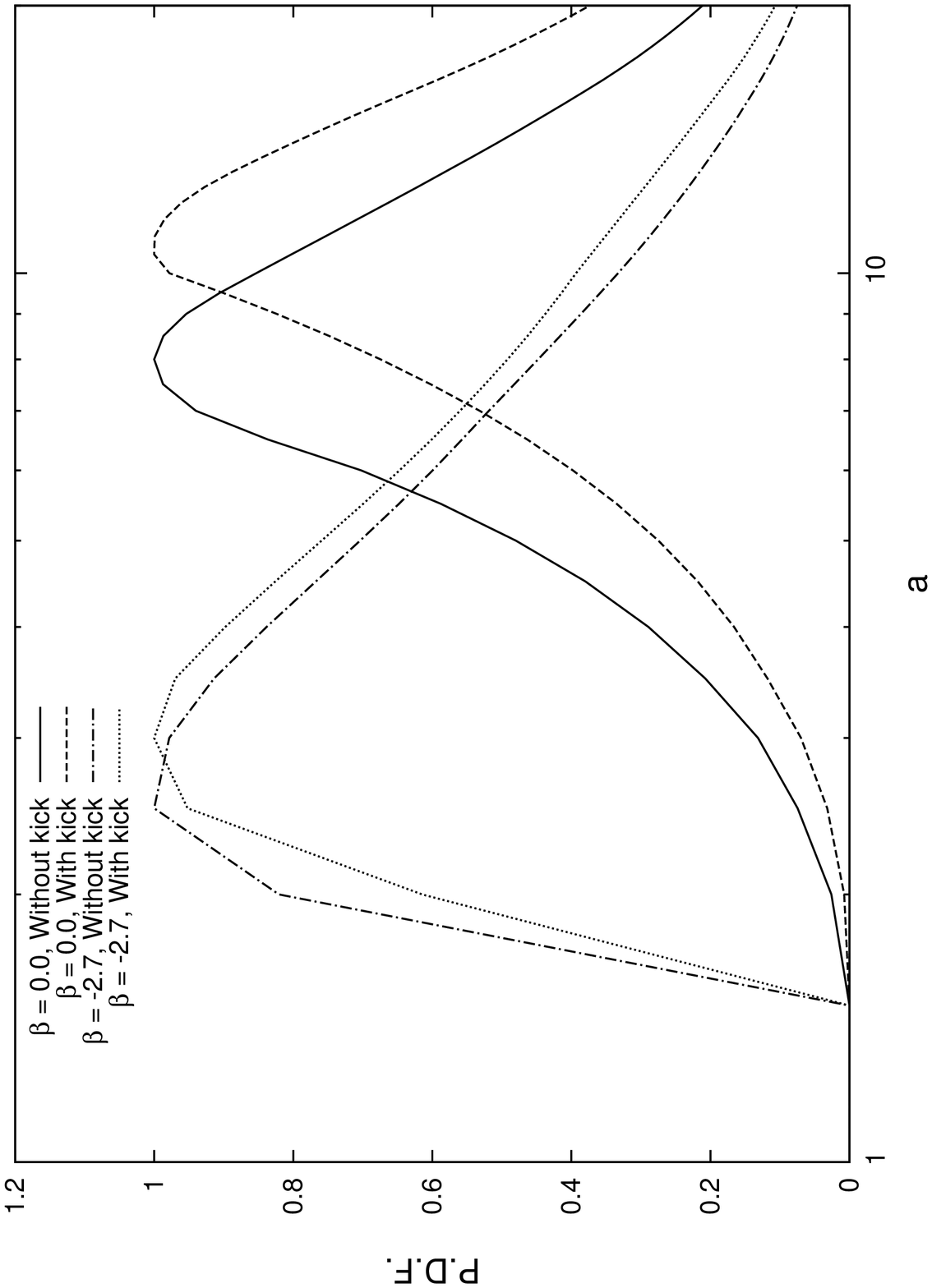}{\it Distribution of the orbital
 separation $a$ for four cases corresponding to two different values
 of $\beta$ and two different SN-kick scenarios. Each case coded by 
line style, as indicated.}{fig:paqpl}

\section{Formation Rate}
\label{frate}

In this section, we describe our recipe for calculating the formation
rate of pre-LMXBs, which serves as an input to the calculation of the 
LMXB X-ray luminosity function (XLF) described in Paper III. 
The first point we notice is that the 
various steps in the formation process of pre-LMXBs described in previous 
sections do not take equal amounts of time. The three timescales that
are of relevance here are  (a) $\tau_{prim}$, the timescale for the primordial 
binary to reach the CE phase, (b) $\tau_{postCE}$, the timescale of evolution of
the post-CE binary upto the SN explosion, and, (c) $\tau_{tid}$, the 
timescale of tidal evolution of the post-SN binary. We first note
that $\tau_{postCE}$, which essentially is the timescale of evolution of a
He-star, is much shorter than the other two timescales. $\tau_{tid}$ can
have a considerable range, depending upon the initial eccentricity of the 
system just after the SN. Typical tidal timescales given in literature are 
$\sim 10^5$ years \citep{zahn, harshal1E}. We assume here that the typical 
range given by these authors works well for the entire relevant range. In 
that case, $\tau_{tid}$ may be an order of magnitude smaller than $\tau_{prim}$,
which is the timescale of the evolution of the primary, $\sim 10^{6-7}$ years. 
Thus the dominant timescale in the evolution of a pre-LMXB is $\tau_{prim}$. 

The formation rate of pre-LMXBs is related to the star formation
rate (SFR). Since the SFR evolves on timescales that are much longer than
$\tau_{prim}$ (typically $\sim 10^9$ years), we can treat the SFR to be
quasistatic during pre-LMXB formation, and say that the formation rate
of pre-LMXBs is roughly equal to the SFR at a (small) timelag of 
$\tau_{prim}$. Since we are ultimately interested in the evolution of these
systems to the LMXB phase, we need to take into account the formation rates
over a long span of time when computing the evolution of pre-LMXBs
into LMXBs. Typical evolutionary timescale of LMXB systems
is $\sim 10^9$ years, so that we need to consider the evolutionary history 
of the SFR over timescales of Gyrs. Evolution of the SFR has been
studied in great detail in the literature over the last fifteen
years or so, using multiwavelength studies of galaxies as well as 
theoretical investigations of the underlying evolutionary processes 
\citep{madau96, madau98, pettini, blain, hartwick}. 
The SFR profiles fitted to the data can be generally divided into two 
classes: \emph{peak}-type and \emph{anvil}-type. (See \citet{blain} 
for a detailed summary of the models in each class). These models 
generally provide the SFR as a function of the redshift $z$, which 
needs to be converted into the lookback time for our purposes here. 
The relation $z \equiv z(t)$ is dependent on the details
of the cosmology assumed: we do not go into the details of  
cosmological models here, but rather assume the standard modern 
prescription in the literature.

We present here a general method of calculating the formation rate of 
pre-LMXBs, applicable to any form of SFR profile. We note first that
$\tau_{prim}$ is roughly equal to the main sequence life time of the primary,
which is a function of the mass of the primary and can be approximated by 
$\tau_{prim} \approx 12 M_p^{-2.5}$ Gyr for the primary mass range of 
interest in this work (See appendix C of \citet{ghoshbook}). 
$M_p$ can be calculated using the inverse transformations, but
we need all three parameters describing the final stage to calculate
$M_p$. The net formation rate of the 
pre-LMXBs at any given time is given by $SFR(t-\tau_{prim})$. The formation
rate for specific values of $M_s$ and $a$ can be calculated by multiplying
the formation rate with the PDF and integrating this product over all
eccentricities, since the net formation rate is dependent 
on eccentricity through $\tau_{prim}$. Thus the 
formation rate as a function of $M_s$ and $a$ can be written as
\begin{equation}
R(M_s, a; t) = \int_0^1 \rm{SFR}(t-\tau_{prim})\,f_{tid}(M_s, a, e)\,de.
\label{eqn:frate}
\end{equation}

The formation rate given by Eq.(\ref{eqn:frate}) can be used as a starting 
point for studying the evolution of LMXBs, which we do in Paper III. As 
noted above, one needs to know $R(M_s, a; t)$ for at least a few Gyr
back from present epoch if one wants to calculate the properties of the 
current population of LMXBs. The importance of cosmic star formation history 
(SFH) on collective properties of LMXBs has been studied for more than a 
decade now \citep{whiteghosh, ghoshwhite}. The work of these authors was a 
first step in this direction, which described the evolution of the total number
of these systems without considering the distributions of the 
system parameters. Our scheme of calculations presented here and in 
Paper III of this series, where we follow the evolution of LMXBs,
enables us to study various collective properties of LMXB populations 
as distributions of luminosity (\ie, XLF) and orbital period.

\section{Discussion}
\label{discuss}

We have described in this paper a method for calculating the formation rate
of pre-LMXBs from given distributions of primordial binaries and given
star formation rates. We studied the pre-LMXB PDF as a function of the 
companion mass and the orbital separation  for various values of CE
parameter, SN-kick scenario, and other parameters. The main
conclusions of this work can be summarised as follows:

\begin{enumerate}

\item LMXB formation is a very tightly constrained process. We showed that
the constraints suggested by \citeauthor{kalogera98} can be
transformed into constraints on primordial binaries and so
demonstrated how only a small allowed region in the phase space of 
primordial binaries is able to produce pre-LMXBs and then possibly
LMXBs (if further conditions are satisfied, \eg, attainment of Roche
lobe contact within a Hubble time). We showed that the CE parameter
was a major factor affecting the allowed phase space, so that a good
understanding of the CE process was essential for modeling the 
collective properties of LMXBs.

\item The PDF of pre-LMXBs was studied in a bivariate form,\ie, 
as a function of $M_c$ and $a$ as well as in monovariate forms for each 
of these variables. These PDFs were shown to agree with the results of
earlier studies in this field. It was shown that a power-law
distribution of the primordial mass ratio $q$ can lead to very
different PDFs for different values of the exponent $\beta$. 
$\beta = 0$ leads to a larger number of wider systems, and a companion 
mass distribution skewed towards the higher end. This would naturally
lead to a larger number of LMXB systems harboring giant companions. 
On the contrary, $\beta = -2.7$ leads to a larger number of compact
systems with smaller-mass companions. This would lead to a larger
number of LMXB systems with main-sequence companions. We take
up these questions in more detail in Paper III.

 \item The effects of the metallicity of the primordial primary and of the
inclusion of natal SN-kicks were also studied.  The former effects are
generally small. For the latter, significant effects come only for
ICCSN-kicks, as the ECSN-kicks are so small as to give results 
essentially identical to those for the no-kick scenario.  

\end{enumerate}

In the next paper (Paper III of the series), we proceed from the 
pre-LMXB formation rate found in this paper to a computation of the
expected LMXB XLF, which we then compare with the observed LMXB XLF
with a view to understanding and constraining the essential processes
of pre-LMXB and LMXB formation and evolution. We re-emphasize that this
effort should be regarded as a proof-of-principle type of exercise 
to understand the basic physics underlying the LMXB XLF, similar to
what we did in Paper I for the HMXB XLF.

\bibliographystyle{apj}
\bibliography{bibfile}

\end{document}